\documentclass[amssymb,aps,superscriptaddress,pre,twocolumn,floatfix]{revtex4-1}
\usepackage[]{amsmath}
\usepackage{graphicx}
\usepackage{epstopdf}
\usepackage{dcolumn}
\usepackage[usenames, dvipsnames]{color}

\definecolor{DarkCyan}{cmyk}{1.0,0.0,0.0, 0.4}
\definecolor{LightRed}{cmyk}{0,0.8,0.9, 0}

\begin{document}

\title{Competing opinions and stubbornness: connecting models to data}

\author{Keith Burghardt}
\email[]{keith@umd.edu}
\affiliation{Dept. Of Physics, University Of Maryland, College Park, Maryland, USA, 20742}

\author{William Rand}
\affiliation{Center for Complexity in Business, University Of Maryland, College Park, Maryland, USA, 20740}
\affiliation{Dept. of Marketing, Robert H. Smith School of Business, University Of Maryland, College Park, Maryland, USA, 20740}

\author{Michelle Girvan}
\affiliation{Dept. Of Physics, University Of Maryland, College Park, Maryland, USA, 20742}
\affiliation{Institute for Physical Science and Technology, University Of Maryland, College Park, Maryland, USA, 20742}
\affiliation{Santa Fe Institute, Santa Fe, New Mexico USA, 87501}

\date{\today}

\begin{abstract}
We introduce a general contagion-like model for competing opinions that includes dynamic resistance to alternative opinions.  We show that this model can describe candidate vote distributions, spatial vote correlations, and a slow approach to opinion consensus with sensible parameter values. These empirical properties of large group dynamics, previously understood using distinct models, may be different aspects of human behavior that can be captured by a more unified model, such as the one introduced in this paper. 

\begin{description}
\item[PACS numbers]
02.50.-r 
87.23.-n 
89.65.-s 


\end{description}

\end{abstract}


\maketitle

\section{Introduction}

The study of opinion dynamics, which has received considerable attention from statistical physicists, network scientists, and social scientists \cite{Sznajd,Contrarian,3Choice,NCO1,VMConsensus,Zealot3,OpDynSS1,OpDynSS2}, explores the dynamics of {\em competing} ideas or opinions via interactions between individuals. Example application areas include voting patterns \cite{VM1,VM2,VMConsensus,VMFreezePeriod,VoterScaling1,VoterScaling2,ZealotVMVoteDist,Zealot1,Zealot2,Zealot3}, product competition \cite{Duopoly}, and the spread of cultural norms and religions \cite{ReligionDynamics,Axelrod,FamilyDist}.
The goal of our work is to gain new insights into opinion dynamics by introducing a well-motivated model that can simultaneously describe multiple empirical observations which have previously been explained by several different models.

A variety of models have been proposed to explain individual features of opinion dynamics observed in empirical data. For example, some models have focused on producing nonconsensus in equilibrium \cite{NCO1,NCO2,Contrarian}, while others can reproduce observed vote distributions \cite{VoterScaling1,ZealotVMVoteDist}, or long-range vote correlations \cite{CorrelationData}. Because we believe these observations are all fundamentally related, we introduce a new model, called the Competing Contagions with Individual Stubbornness (CCIS) model, which can robustly explain the above behaviors using agent-based dynamics designed to mimic observed human behaviors. Not only does the CCIS model match the aforementioned observations with consistent parameter values, it is general enough to incorporate a wide array of plausible factors affecting the success of opinions in the real world, allowing for agents with a neutral state, opinions that are stronger than others, and opinions that may be introduced after an earlier opinion has spread through a population. Here, for simplified modeling and analysis, we focus on the case of equal strength opinions introduced at the same time and leave these other cases for future work.

In the CCIS model, at any given time point, individuals can either be in a neutral state or in one of $Q$ different opinion states. Opinions can change over time as individuals try to ``convince" others in their social network to adopt their opinion. In our model, individuals exhibit "stubbornness," meaning that the longer an opinionated individual keeps his or her opinion, the less likely they are to switch to a new one. This property has been seen empirically in previous studies \cite{PTP}. We distinguish this from other models in which individuals resist changes in their opinion independent of time, e.g., \cite{Zealot1,Zealot2,Zealot3,ZealotVMVoteDist,ResistanceGalam1,ResistanceGalam2}. Within the CCIS model, individuals that have held on to their opinion for a long time will eventually completely lose the ability to be convinced by one of their neighbors to adopt a different opinion.  However, all opinionated individuals move back to the neutral state at a constant rate, which is designed to allow for a large fraction of  ``independent" voters, as is the case for the United States electorate \cite{PewResearch}. Once an individual becomes neutral, they can switch opinions to any of their neighbors', which creates longer timescale opinion dynamics. 


The remainder of the paper is structured as follows. We first describe related work (Section \ref{RelatedWork}) and then provide the details of our model and algorithm implementation (Section \ref{ModelDetails}), before comparing the results of our model to empirical data (Section \ref{DataAgreement}). We then analyze the dynamics of our model using a series of approximations (Section \ref{Analysis}) and numerically study the consensus time outside of the parameter ranges for which our analysis is valid (Section \ref{ConsensusTimes}). Finally, we conclude with a discussion of future work (Section \ref{Conclusion}).

\section{Related Work}
\label{RelatedWork}

In this section, we review the empirical studies that motivate the CCIS model and we discuss related models.

In recent years, large sets of empirical data have allowed researchers to better observe collective social dynamics \cite{Facebook,ComplexContagion,LermanCC,CorrelationData,PollVolatility,
VoterScaling1,VoterScaling2}, leading to new insights in the field. We first focus on two themes that have received recent attention: candidate vote distributions \cite{CorrelationData,VoterScaling1,VoterScaling2} and spatial vote correlations \cite{CorrelationData}.

Two important studies on election data from several countries demonstrate that vote distributions, when rescaled by $Q/N$, where $Q$ is the number of candidates and $N$ is the number of voters, often collapse to a universal distribution (see inset of Fig. \ref{QScaling}) \cite{VoterScaling1,VoterScaling2}. Two recent models have been proposed to explain this behavior \cite{VoterScaling1,ZealotVMVoteDist}. 

A model by Fortunato and Castellano \cite{VoterScaling1} assumes that voters are convinced to vote for a specific candidate unique to each of $Q$ social networks, with no interaction between voters of opposing candidates. While the model provides good agreement with vote distribution data and demonstrates how ``word-of-mouth'' or contagion-style spreading can play an important role in observed voting patterns, it cannot capture one important feature of real elections $-$ that candidates seem to often compete for a common set of voters \cite{MedianVoter1,MedianVoter2,MedianVoter3}. Hence, we believe that a model with competing opinions on a single network, such as the one introduced in this paper, is needed to for a more complete picture of how individual level dynamics can translate to observed voting patterns.

Another model by Palombi and Toti, which does include interactions between supporters of different candidates, yields qualitative agreement with empirical data on vote distributions  by assuming a network of interactions with significant structure (non-overlapping cliques connected by sparse random links) as well as a distribution of zealots (unwavering candidate supporters) that is related to the underlying clique structure of the network. By contrast, our goal is to show agreement with empirical data on both vote distributions and voter correlations using a somewhat more generic network of interactions and without imposing any connection between candidate preferences and network placement for any individuals. The contagion-inspired framework of our CCIS model, e.g., the inclusion of a neutral state and a tunable transmissability parameter, gives it the flexibility to match the two aforementioned empirical patterns of interest while simultaneously remaining relatively simple.


Recent empirical studies have shown that the spatial correlation of vote-shares in United States elections and the spatial correlation of turn-out rates in European elections decreases as the log of the distance between two voting districts \cite{CorrelationData,CorrelationData2}. This contrasts to correlations of spins in many statistical mechanic spin models, which decrease as a power law or exponentially with distance \cite{Onsager}, but is a prediction of some spin (or opinion) models, such as the VM, at an arbitrary, fixed time \cite{VM1,VM2,CorrelationTheory}. 


In addition to matching these empirical patterns by yielding spatial opinion correlations that decrease as the log of the distance between individuals (in the case of networks with significant spatial structure), the CCIS model shares other important features with the well-studied VM. In the VM, at each time step, an individual chooses to adopt the opinion of one of their randomly chosen neighbors \cite{VM1,VM2}. In the basic CCIS model, opinions also change via interactions with neighbors, but instead of interacting with one neighbor at a time, individuals try to persuade all their neighbors simultaneously, similar to the approach used in the aforementioned Fortunato and Castellano~\cite{VoterScaling1} paper. In Section \ref{Analysis}, we also consider CCIS type dynamics for the situation in which, as in the VM, interactions at each time step are focused on an pair of connected individuals instead of one individual and all of their neighbors.


The CCIS model also has important similarities to the well-studied Susceptible-Infected-Susceptible (SIS) model from epidemics. In the SIS model, individuals exist in only one of two states: ``susceptible'' and ``infected,'' and infections propagate via contacts between infected and susceptible individuals, with infected individuals eventually recovering to the susceptible state. The SIS model can be applied to the study of opinion dynamics, but, because the basic model is an explicitly a two-state model, it can only be used to explore how a single opinion (contagion strain) propagates through a neutral (susceptible) population, and the SIS model must be modified to explore the competition dynamics among multiple opinions. 

A few recent studies have modeled the coexistence of two contagion strains on networks with SIS-like models \cite{PathogenCompetitionWTA,PathogenCompetition2,PathogenCompetition1,
PathogenCompetition3,EvolutInfect,2StrainFinitePopulation,Mutation}.  
Typically, in these models, individuals can only switch from one strain to another if they recover first \cite{EvolutInfect,2StrainFinitePopulation,PathogenCompetition2}, or else two strains can cohabit a single individual but interact on coupled networks \cite{PathogenCompetition3}.
In the CCIS model, however, individuals can switch directly between opinions instead of first moving to the recovered state, and all opinions propagate on a single network. Furthermore, no individual can have more than one opinion at any time. These are realistic assumptions for opinion dynamics, because individuals can directly switch between opinions more easily than they might directly switch between diseases, and would be unlikely to hold contrasting opinions at the same time. We note, however, that across a wide parameter space in our model, one opinion eventually dominates (e.g., Eq. \ref{DiffusiveConsensus} and Figs. 7, 8, \& 9), while the contagion models described above have large parameter regimes where two contagions can stably coexist. In Section \ref{Analysis}, we discuss in more detail how the CCIS model approaches consensus. 

The CCIS model is further distinguished from the VM and SIS model by having individuals exhibit stubbornness \cite{FastConsensus} (similar model assumptions are made in other works \cite{VMFreezePeriod,NoiseReducedVM,GalamBuildInflex,CODA}). In our definition of stubbornness, individuals increasingly resist changing their opinion, in contrast to other models where individuals resist changes in their opinion independent of time, e.g., \cite{Zealot1,Zealot2,Zealot3,ZealotVMVoteDist,ResistanceGalam1,ResistanceGalam2}. In pre-trial publicity (PTP) experiments \cite{PTP}, the correlation between the jury decision and the PTP opinion is stronger when individuals are exposed
to PTP more than a week before the mock trial compared to when the exposure happens closer to the start of the trial.
 This provides some evidence that individuals change their resistance to alternative opinions, but not necessarily monotonically with time. Further evidence from voter data is currently lacking and is an important area for future study. Nonetheless, the initial evidence from jury experiments and the strong agreement to data we find with our current model is suggestive that stubbornness may play an important role in the dynamics of opinions.
We also note that stubbornness is similar to the {\em primacy effect}, well studied in psychology \cite{Primacy1,Primacy3}, in which the first idea someone hears is favored regardless of its validity. That effect, however, deals only with the ordering of choices and does not take into account the time intervals between choices. 

The CCIS model is designed to offer a more general framework than many previous models. It allows for different opinions to be more or less likely to be adopted relative to each other, for individual opinions to be more or less likely to exhibit stubbornness, for some opinions to be introduced at later times than others, and for individuals to exist in a neutral state. These additions give it the flexibility to capture a variety of situations. In this paper, for simplicity, we focus on the case of opinions with equal strengths and individuals with identical stubbornness parameters.

\section{Model Details}
\label{ModelDetails}

\begin{figure}[tbp]
\includegraphics[scale=0.48]{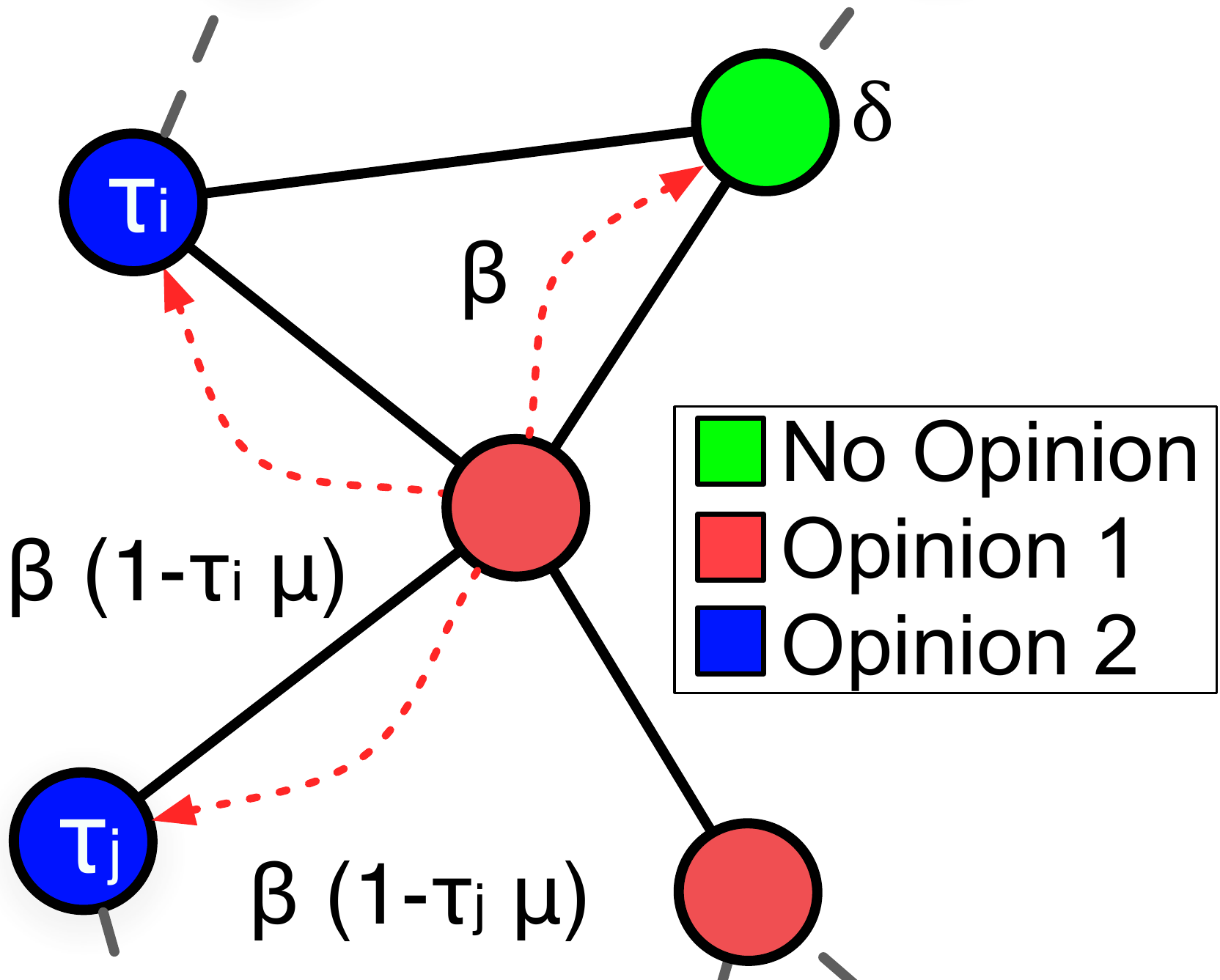}
\caption{\label{ModelSchematic}(Color online) The schematic of our model. Arrows indicate attempts to convince neighboring individuals, with probabilities for success appearing next to each arrow. The length of time the nodes have held their current opinion is indicated by the text inside the node.}
\end{figure}

\begin{figure}[tbp]
\includegraphics[scale=0.325]{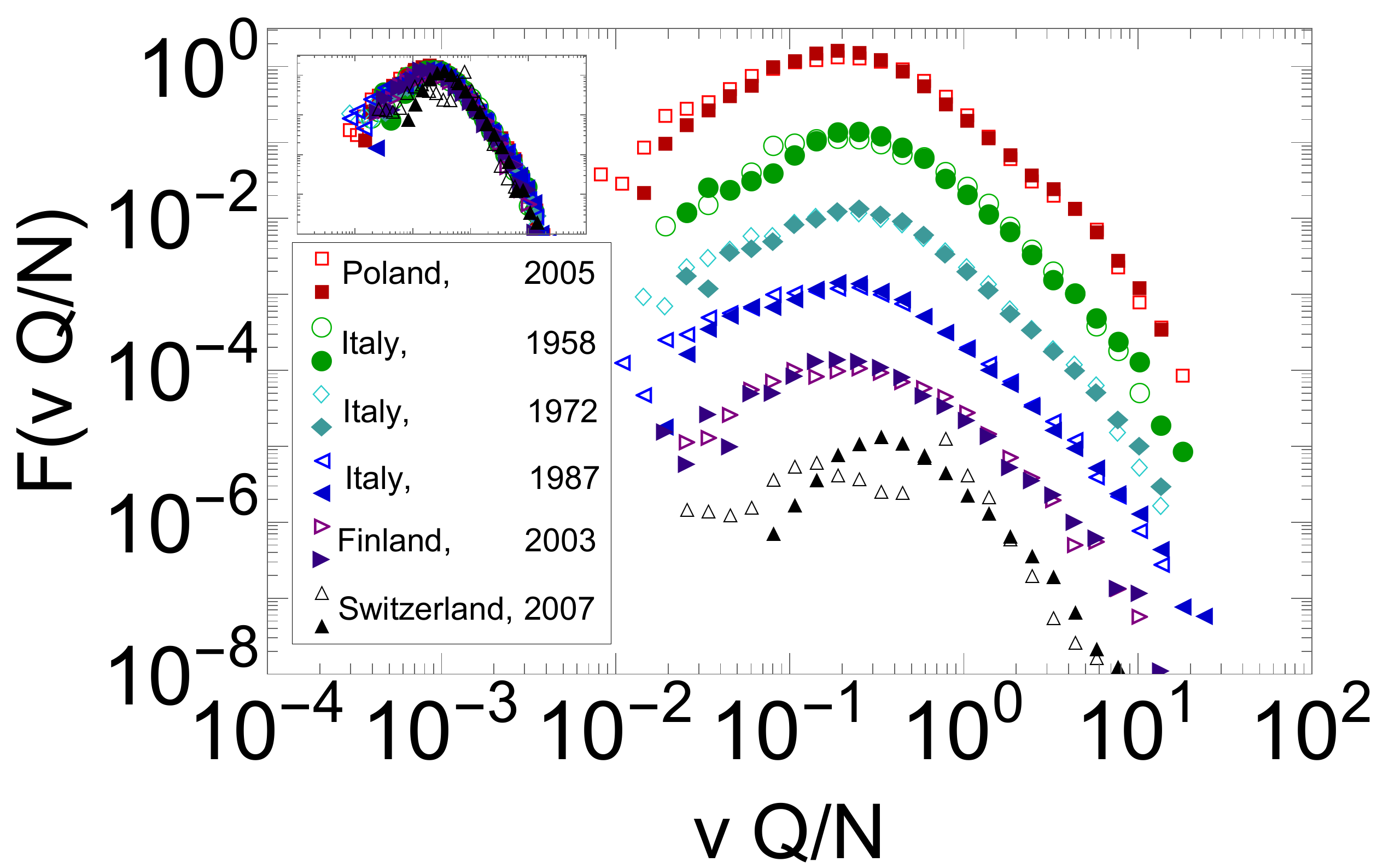}
\caption{\label{QScaling}(Color online) A comparison of scaled vote distributions between the CCIS model (closed markers) and elections (open markers) (data from \cite{VoterScaling2}), in which data is shifted down by decades for clarity (inset shows the original data collapse). Here $v$ corresponds to the number of votes, with the number of candidates, $Q$, and  size of the population, $N$, equal to the empirical data values. The initial fraction seeded with a preference to a candidate is fitted to the scaled vote distribution of Poland's 2005 elections by Maximum Likelihood Estimation. All other parameters are fixed.}
\end{figure}

In this section, we describe the dynamics of the CCIS model in detail (see Fig. \ref{ModelSchematic} for a schematic).
The model operates on a network with $N$ nodes, in which the state of each node, $i$, is $s_i \in \{0, 1,2,...,Q\}$, where $Q$ is the total number of opinionated states and 0 corresponds to the neutral state. For ease of analysis, we study the case in which interactions between individuals occur on a fixed, unweighted network.

At time $t = 0$, $n_0$ (possibly 0) nodes are in state 0, $n_1$ (again, possibly 0) are in state 1, etc., such that $n_0 + n_1 + ... + n_Q =N$. We leave open the possibility for new opinions to be added at arbitrary times in the simulation. However, in this paper, we focus on the case where at $t=0$, $n_1 = n_2 =...=n_Q$ (and therefore all opinions are simultaneously introduced).

Algorithmically, we implement the model as follows:

\begin{enumerate}
\item Pick a random opinionated node $i$ (i.e, a node not in state 0) 
	\begin{enumerate}
		\item Revert $i$'s state to 0 with probability $\frac{\delta}{1+\delta}$
		\item Otherwise pick each of $i$'s neighbor at random:
		\begin{enumerate}
				\item Convert any neutral (state 0) neighbor to state $s_i$ with probability $\beta$
				\item Convert any contrary opinionated neighbor $j$ to state $s_i$ with probability max$\{\beta(1-\tau_j \mu),0\}$, where $\tau_j$ is the time since node $j$ adopted its current opinion.
	\end{enumerate}
	\end{enumerate}
\item Count the number of opinionated individuals, $N_\text{op} = N - n_0$, and repeat from step 1 with time incremented by $\Delta t = N_\text{op}(1+\delta)^{-1}$.
\end{enumerate}
Here, for simplicity, we assume that the persuasiveness of each individual, $\beta$, the recovery rate, $\delta$, and the stubbornness, $\mu$, does not depend on which opinion is held, but there may be situations for which these parameters should be differentiated according to opinion. We implement stubbornness in the following way: the effective persuadability of a node $j$ by a neighbor with a contrary opinion, $\beta(1-\tau_j \mu)$, decreases linearly in time until $\tau_j = \mu^{-1}$, at which point individual $j$'s opinion remains fixed unless $j$ moves to the neutral state, which occurs at rate $\delta$. A natural alternative to our implementation of stubbornness is to construct an effective persuadability that decreases exponentially, $\beta \text{exp}(-\tau_i\mu)$. We choose the linear form for its simplicity, but we expect similar dynamics for the two cases.

Table I summarizes that model's parameters and variables.

\begin{flushleft}
\begin{minipage}{\linewidth}
\label{Symbols}
\centering
Table I: Definitions of Symbols and Parameters\\

\begin{tabular}{l l}
  \hline\hline
\vspace{-10pt}\\
  \textbf{Symbol} & ~~~~~~~~~~~~~~~~~~~~\textbf{Definition} \\
\hline
\vspace{-9pt}\\
$t$ & Time \\
$\tau$ & Time the most recent opinion has been kept \\
$\beta$ & Persuasiveness \\
$\mu$ & Stubbornness rate\\
$\delta$ & Recovery rate \\
$Q$ & Number of opinions\\
$N$ & Number of nodes (``voters")\\
$\rho^{(A)}$ & Opinion $A$ density as a function of $t$ and $\tau$\\
$P^{(A)}(t)$ & $\int_0^\infty \rho^{(A)} (t,\tau') d\tau'$\\
$\alpha$ & Scale-free degree distribution coefficient\\ & ($p(k)\sim k^{-\alpha}$)\\
\hline
\hline
\end{tabular}
\end{minipage}
\end{flushleft}

Note that at each time step, $\Delta t$, is normalized such that $N_\text{op}$ node-node interactions take place, and $\delta N_\text{op}$ of the opinionated nodes recover, after a time $\sum_i \Delta t_i = 1$. Holding $N_\text{op}$ constant for each time step, $\Delta t = ((1+\delta) N_\text{op})^{-1}$ and the recovery probability is $\delta/(1+\delta)$. This method is based upon a similar approached used for the SIS model to approximate continuous time dynamics \cite{EpidThreshold}.

We include the recovery rate in our model to allow for a large fraction of individuals to remain neutral over long time scales.
This is motivated in part by the empirical observation that a significant fraction of Americans remain unaffiliated with any political party, and that this fraction is stable over the timescale of years \cite{PewResearch}, yet in individual elections, these ``independents'' frequently vote for candidates with party affiliations, and hence can be thought of as having adopted the party ``opinion'' over short timescales.   
Additional elements of realism, such as mass media \cite{MassMedia}, party affiliation \cite{HeteroVM}, and variations in the recovery rate, have been left out of this model for simplicity, and may be important for future study.

\section{Agreement With Data}
\label{DataAgreement}

In this section, we show that the CCIS model can reproduce two empirical observations: (1) distributions of votes received by candidates, when appropriately rescaled, follow a nearly universal function \cite{VoterScaling1,VoterScaling2} and (2) correlations between voters decrease only logarithmically as a function of distance \cite{CorrelationData,CorrelationData2}. We find agreement between the CCIS model and both empirical observations using spatially extended networks with heavy-tailed degree distribution  (a reasonable model for social networks \cite{BA,Smallworld}). In agreement with Fortunato and Castellano~\cite{VoterScaling1}, we find that a heavy-tailed degree distribution is important for matching the opinion model's distribution to the empirical vote distribution data. We emphasize that the spatial component (meaning that nodes preferentially connect to others that are spatially close) is necessary to create spatial correlations that match empirical observation. The networks are created as follows: all nodes are embedded on an $\sqrt{N} \times \sqrt{N}$ two-dimensional grid with periodic boundary conditions. The out-degree, $k_i \ge k_\text{min}$, is chosen from from a power law degree distribution, $p(k) \sim k^{-\alpha}$ with minimum degree $k_\text{min}$, which is specified so that the desired average degree, $\langle k\rangle$, is reached. Directed links from node $i$ to the  $k_i$ nearest (in grid-space) other nodes are then created. A fraction $f$ of edges are then rewired at random to add noise to the network. A more detailed description of the network is given in Appendix \ref{fitting}.

\begin{figure}[tbp]
\includegraphics[scale=0.5]{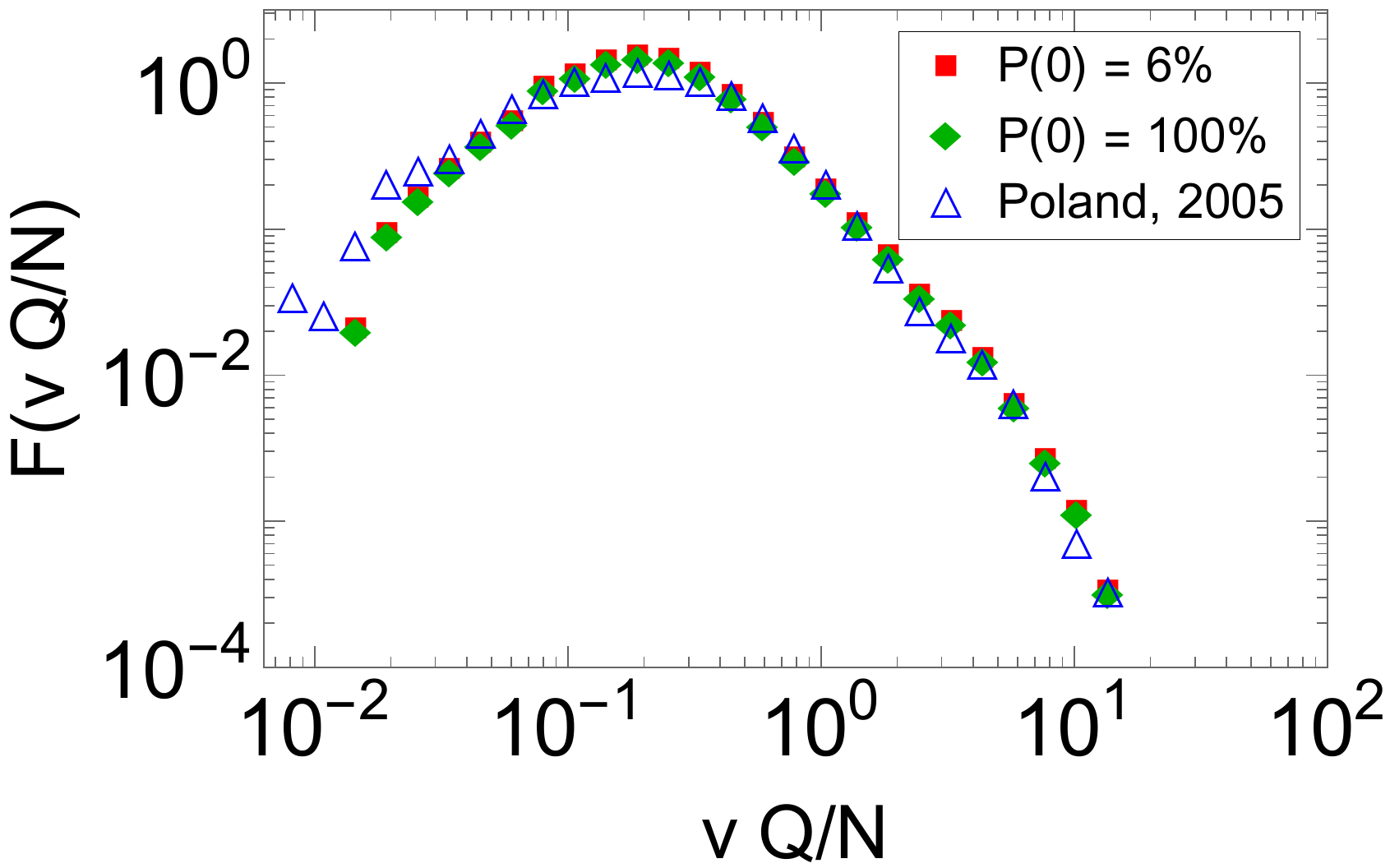}
\caption{\label{InitInfComparison}(Color online) A comparison of the best fits between the CCIS model and the 2005 Poland elections with a fraction of nodes initially seeded with an opinion ($P(0)$) equal to $6\%$ and $100\%$ (see Table I for the definitions of parameters). The parameters are the same except when $P(0) = 6\%$, $\beta = 0.1$, and when $P(0) = 100\%$, $\beta = 0.65$.
}
\end{figure}

\subsection{Voter Scaling}

As Fig. \ref{QScaling} shows, the CCIS model with appropriate parameter choices can closely match empirical vote distributions rescaled by $Q/N$. We simulate each election one time for each set of parameters to test how well our model can typically follow the empirical data, and each election is run on a spatially distributed scale-free graph (as described above) with $N$ and $Q$ the same as empirical data to account for finite size effects. We vary the initial fraction of individuals seeded until the model fits the distribution from Poland's 2005 elections (which has the largest number of elections). All other simulation parameters are fixed to reasonable values: $\beta = 0.1$, $\langle k\rangle = 10$, $\mu = 1$, $\delta = 0$, and $\alpha=2.01$ (see Appendix \ref{fitting} for details regarding the fit and the robustness of the results to changes in the parameters).

\begin{figure}[tbp]
\includegraphics[scale=0.495]{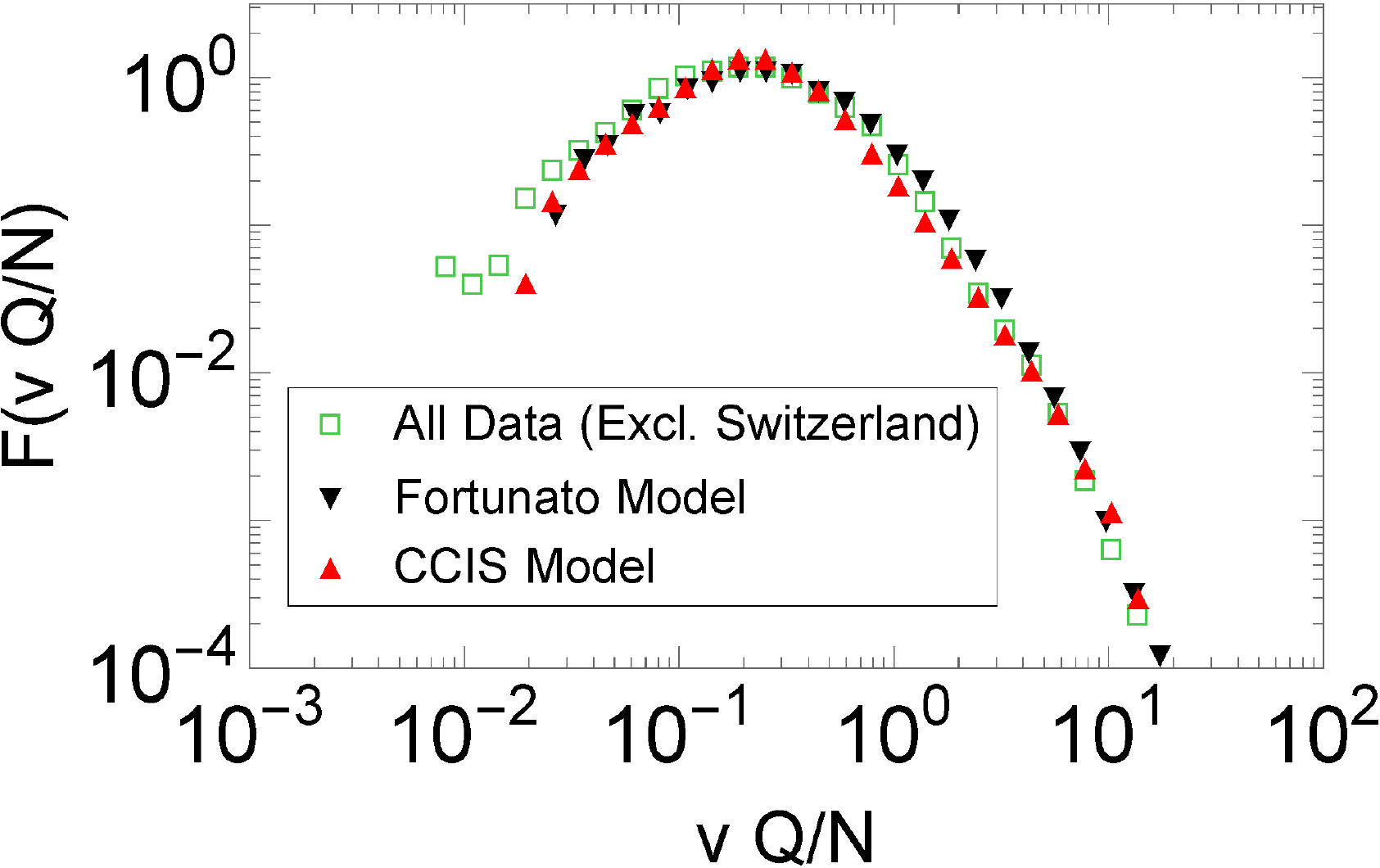}
\caption{\label{Comparison}(Color online) We plot the distribution across all elections shown in Fig. \ref{QScaling}, excluding Switzerland, and compare our fit to the fit of the FC model \cite{VoterScaling1}.
}
\end{figure}

The simulation results plotted are for networks without random rewiring (i.e., $f=0$), but we find similarly good fits for larger values of $f$.
In the simulations, $\mu > 0$ and $\delta = 0$ in order to reach a non-consensus equilibrium, because otherwise we would have to stop the simulation at some arbitrary time before consensus is reached. These same parameters were used to fit all the other countries' elections. 

Overall, we find good fits between our model and voter data as long as $\mu>0$, and the distribution is sufficiently heavy tailed, i.e. the magnitude of the degree distribution exponent is small ($\alpha < 3$).  See Fig. \ref{LE} in Appendix \ref{fitting} for a detailed analysis of the robustness of the fit to parameter variation. Our findings suggest that both individual stubbornness and heavy-tailed degree distributions in social networks \cite{BA} may be important underlying drivers of the generic behaviors observed in opinion dynamics.

The reason for the strong fit in Fig. \ref{QScaling} is in part because our model appears to follow a nearly universal distribution when each vote is rescaled by $Q/N$, like the empirical data from the elections it attempts to model. 
Of the elections modeled, we find that only Switzerland's diverges significantly from our model due to its unusual ``double-hump" distribution, plausibly because votes are swayed by the local language differences (primarily French and German). 

\begin{figure}[tbp]
\includegraphics[width=0.45\textwidth]{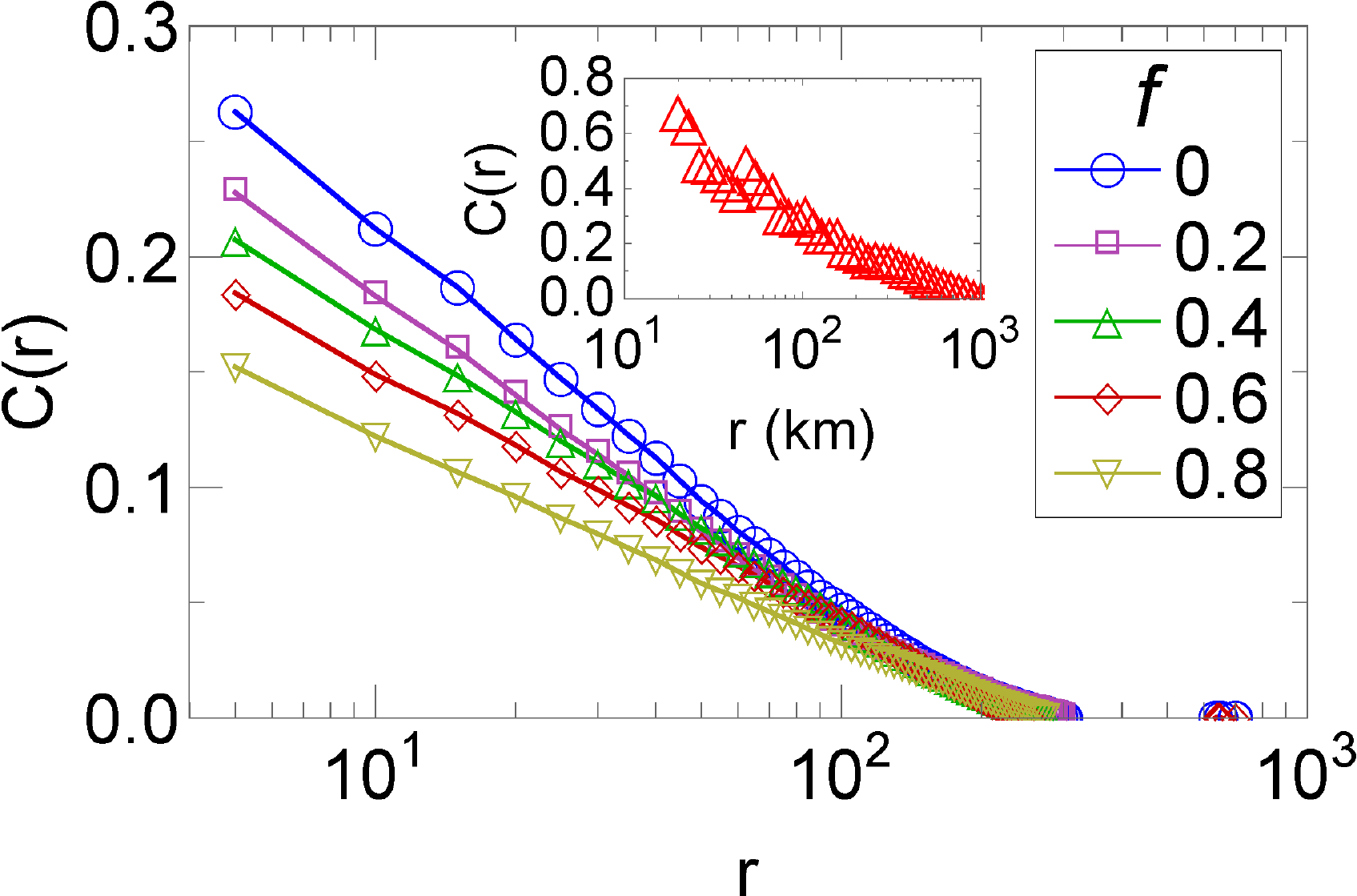}
\caption{\label{SpatialCorrel}(Color online) The correlation as a function of distance for the CCIS model (where nodes are separated by a unit 1 distance on a $10^6$ node network). The CCIS model parameters are the same as in Fig. \ref{QScaling} except here $Q = 2$ and $100\%$ of nodes seeded. $f$ fraction of edges are randomly rewired on a scale-free spatially distributed graph ($f=0$ corresponds to the network in Fig. \ref{QScaling}) showing that the logarithmically decreasing correlations are robust. Inset: similar correlations are seen for data from the year 2000 United States Presidential election \cite{CorrelationData2}.}
\end{figure}

Agreement between the model and empirical data (Fig. \ref{QScaling}) is also possible when the initial fraction of individuals seeded, $P(t=0)$, is $100\%$ if the persuasiveness of each individual, $\beta$, is adjusted to 0.65 (see Fig. \ref{InitInfComparison}). In this case, because $\delta = 0$, no individual ever reaches the neutral state. Despite the fact that agreement with data can be achieved without the inclusion of a neutral state, we believe that 
such a state is important because most voters start out with little knowledge of the candidates. 

One natural way to seed opinions when explaining the candidate vote distribution is to assume that only one individual has an initial vote preference: the candidate himself. This creates a poor fit for our model (not shown), possibly suggesting that the initial spreading process differs from the one that takes over after a short time.


Our work is influenced by the Fortunato and Castellano (FC) model (introduced in Section \ref{RelatedWork}), which was developed to describe the same distribution data \cite{VoterScaling1}.  In both the FC and CCIS model, individuals try to persuade neutral neighbors in the network at some rate. Opinions do not compete in the FC model, but instead spread within isolated networks, meaning that each of the $Q$ candidates convince voters to vote for him or her by word of mouth to their friends, which then  spreads to their friends' friends, etc. In this scenario, an individual only decides whether or not to vote for one specific candidate and never decides between candidates.  
The CCIS model is designed to capture a more realistic scenario in which candidates compete for the same set of voters \cite{MedianVoter1,MedianVoter2,MedianVoter3}.
We directly compare our model to the FC model in Fig. \ref{Comparison}. Both models create similar fits, based on the log-likelihood function, with neither being significantly better.

\subsection{Spatial Correlation}

Next, we show that the CCIS model creates correlations that decrease logarithmically with distance, as seen in empirical studies \cite{CorrelationData,CorrelationData2}. This behavior is not unique to our model because many models can create logarithmically decreasing correlations as they approach the VM Universality Class \cite{VMUniversalityClass} in some special parameter range. We find it important, however, that our model is the first model we are aware of that can reproduce both the previously mentioned vote distributions, and this behavior, especially over a wide set of parameters. In comparison, the FC model \cite{VoterScaling1} assumes non-interacting opinions on random graphs, and the Palombi and Toti model \cite{ZealotVMVoteDist} assumes opinions interact on non-spatially distributed cliques with edges connected randomly between them, so votes are uncorrelated in space. 
Analysis of the observed logarithmic correlations in the CCIS model are discussed in the next section. Simulations, however, suggest the most important property in our model to reproduce the empirical observations is a spatial structure in our social network, whether the network is a lattice, small-world (random rewiring), or the current scale-free spatial network. Therefore, this property is very general, and should be generically seen in empirical data.

Figure \ref{SpatialCorrel} shows results from simulations of our model on spatial scale-free networks with $10^6$ nodes and the same model parameters as in Fig. \ref{QScaling} (if $f = 0$). The figure also shows results from simulations for which a fraction, $f$, of edges were randomly rewired. The rewiring process reduces the spatial features of the graph by creating long-range ties that significantly reduce the mean geodesic distance between points. Even with large $f$, however, we still see strong qualitative agreement with empirical data. 

We note, however, that while empirical voting patterns are consistent with the CCIS model operating on a spatially-extended network, we cannot rule out the possibility that the empirical correlation data is the result of self-segregation, e.g., that ``Republicans" move to ``Republican" counties. Additional data is necessary to differentiate these two potential explanations for spatial correlations in voting behavior.

\section{Analysis}
\label{Analysis}

In this section, we analyze the dynamics of our model to better understand the behaviors it is capable of producing.
To do so, we simplify the model in three different ways, allowing us to probe the dynamics more thoroughly than any single approximation. 

First, to probe the spatial correlation behavior discussed in the previous section we explore the limit in which our model simplifies to a diffusion process.
  Second, we explain how opinion sizes change in time with a transport-like equation, which assumes individuals mix homogeneously in an infinitely large network and tracks the time evolution of the density of individuals who have held a specified opinion for designated length of time. Finally, we use the Fokker-Planck equation to explore, for the case $\mu = \delta = 0$ (i.e., no stubbornness and no recovery), how our model reaches opinion consensus for finite systems with heterogeneity in the connectedness of individuals. Under the Fokker-Planck approximation (FPA), we handle heterogeneity in the number of connections but we do not capture spatial effects or incorporate stubbornness and recovery, motivating all three separate types of analysis.

\subsection{Spatial Correlations}

Spatial correlations between opinions in the CCIS model decrease logarithmically over a wide parameter space (see Fig. \ref{SpatialCorrel}). We can demonstrate this spatial correlation behavior analytically for the continuum limit of the CCIS model seeded with two opinions (and no neutral individuals) on a lattice grid, for the case $\mu = \delta = 0$.
Because $\delta = 0$, nodes do not independently change to any other state, and furthermore, because $\mu = 0$, the probability of each node changing their state is $(number~of~opposing~neighbors)/[(2 d)^2 \beta]$ at any timestep, where $2 d$ is the degree of a $d$-dimensional lattice.
In comparison, the two-opinion VM assumes that agents are convinced by a random neighbor's opinion at each timestep \cite{VM1,VM2}, or equivalently, the probability of any node changing their state is $(number~of~opposing~neighbors)/(2 d)$, therefore, in this parameter range, the CCIS kinetics is exactly the same as the VM, with time scaled by $2 d \beta$.

\begin{figure}[tbp]
\includegraphics[scale=0.8]{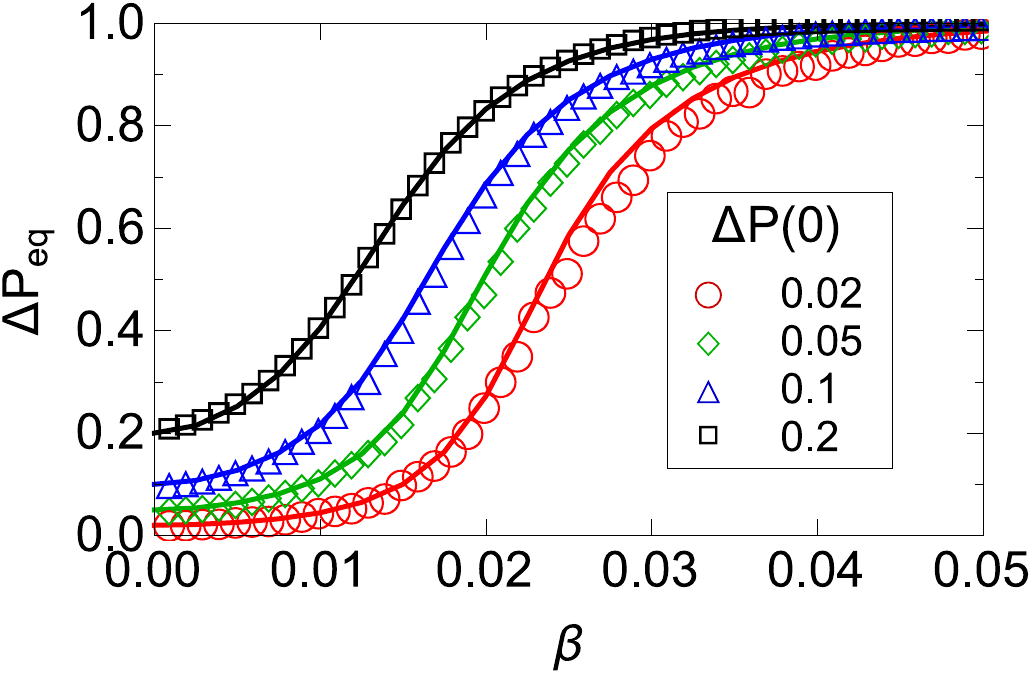}
\caption{\label{TheoryVsSim}(Color online) The difference in equilibrium opinion densities, $\Delta P \equiv |P^{(1)} - P^{(2)}|$, as a function of $\beta$ between theory (solid lines) and simulations, where $\delta = 0$ and $\mu = 0.2$. $\Delta P = 0$ corresponds to a 50/50 split in opinions while $\Delta P = 1$ corresponds to complete consensus. Simulations are on networks $N = 10^5$ and degree $k = 10^2$.}
\end{figure}

 The VM can be approximated as a diffusion process in the continuum limit \cite{CorrelationTheory}, meaning the correlation as a function of time, $t$, can be expressed as:
\begin{equation}
C(r) \sim 
\begin{cases}
1 - \frac{r}{\sqrt{D t}} & d =1 \\
 1 - \frac{\text{log}(r)}{\text{log}(\sqrt{D t})} & d = 2 \\

 r^{2-d} & d \ge 3 
\end{cases},
\label{Correl}
 \end{equation}
\vspace{6pt}

\noindent
in which $D = d$, $r$ is the distance between nodes, and nodes are separated from their neighbors by a distance of one unit.  Eq. \ref{Correl} is the same for the CCIS model in this limit, with $D = (2 d^2 \beta)^{-1}$ to reflect the rescaling of time. The spatial correlation between opinions in the CCIS model therefore decreases as log$(r)$ for fixed time in this limit.  


\subsection{Transport-Like Approximation (TLA)}

Next, we try to better understand how opinions change in time in the CCIS model. We present a partial differential equation similar to the transport equation, to describe the dynamics of the CCIS model in the mean field. This approximation, which we discuss in more detail in Appendix \ref{TLA}, holds for all $\beta$, $\mu>0$, and $\delta = 0$:

\begin{widetext}
\begin{equation}
\label{MFT}
(\partial_t + \partial_{\tau} ) \rho^{(A)}(t,\tau) = 
- \Theta(1-\tau \mu) (1 - \tau\mu)\beta k\rho^{(A)}(t,\tau)\sum_{B\neq A} P^{(B)}(t).
\end{equation}
\end{widetext}

Here, $\rho^{(A)}(t,\tau)$ is the density of individuals at time $t$ that have have opinion $A$ for a time $\tau$. The above equation says that $\rho^{(A)}(t,\tau) \rightarrow \rho^{(A)}(t+\Delta t,\tau + \Delta t)$, 
 and change to an opinion $B\neq A$ at a rate $\beta (1-\tau\mu)$.
If $\tau \mu > 1$, the RHS is 0 due to the Heaviside step function, $\Theta$. The boundary condition (not shown) describes the gain in new individuals (increase in $\rho^{(A)}(t,0)$) via conversion of individuals who were neutral or of an opposing opinion, allowing $P^{(A)}(t) = \int\rho^{(A)}(t,\tau')d\tau'$ to remain constant in equilibrium. Agreement between the equation and simulations is poor when $\mu = 0$, because, after being stochastically pushed out of equilibrium, the system quickly approaches consensus. Similar results are seen when $\delta > 0$, after incorporating a few additional terms. We will discuss how to analyse the dynamics when $\delta > 0$ in the next section. 
However, excellent agreement between theory and simulations is observed in Fig. \ref{TheoryVsSim} when $\delta = 0$ and $\mu > 0$.

\subsection{Fokker-Planck Approximation of the CCIS Model}

We can also analyze the model when $\mu = \delta = 0$, with the Fokker-Plank Approximation (FPA). The main difference between the FPA and the TLA is that the FPA takes into account the size of the system, and degree heterogeneity of a random graph, but does not incorporate the effects of stubbornness or recovery. Under this approximation, links randomly rewire, so we have no spatial information about the network, and cannot say anything about spatial correlations. It is therefore a powerful theory but only for specific network topologies. Our analysis may be improved upon, by modeling bipartite networks, networks with strong cliques, or using a more accurate pair approximation \cite{VMConsensus,PAHomo,PAHetero}, but our goal here is to derive simple expressions that can describe some of the most interesting behavior. We give the details of the FPA in Appendix \ref{FPA} and describe the main results here.

Consensus time, $T_\text{cons}$, is found to be finite and scales in non-trivial ways with the network topology and the persuasiveness parameter $\beta$. If $\rho$ are the fraction of individuals with one of two opinions, we find that
\begin{equation}
\label{FPEqu}
\frac{\rho (1-\rho)}{N_\text{eff}}\frac{\partial^2 T_\text{cons}}{\partial \rho^2}  = -1,
\end{equation}
where $N_\text{eff}$ is the effective size of the network:
\begin{equation}
\label{DiffusiveConsensus}
N_\text{eff} = 
\begin{cases}
\frac{N}{\beta^2 \langle k^2 \rangle} & \text{Outward Process} \\
\frac{N}{\beta^2 \langle k \rangle^2} & \text{Neutral Process} \\
\frac{N}{\beta^2 \langle k^2 \rangle} & \text{Inward Process} \\
\end{cases},
\end{equation}

\noindent
and where $\langle k\rangle$ and $\langle k^2\rangle$ are the first and second moments, respectively, of the network degree distribution. 
Solving Eq. \ref{FPEqu}, we find that $T_\text{cons} \sim N_\text{eff}$ (see Appendix \ref{FPA} for derivation).

In Eq. \ref{DiffusiveConsensus}, the outward process is where an opinion spreads from an individual to its neighbors (which is assumed in the basic CCIS model). More generally, there are two other ways the opinion could spread: (1) the neutral process is where an opinion spreads between two individuals on a random link, and (2) the inward process is where opinions spread from neighbors to an individual.

We now discuss comparisons between simulations and theory for the outward process (in Appendix \ref{FPA}, we compare $T_\text{cons}$ in simulations to an equivalent $T_\text{cons}$ theory for the neutral and inward processes).

\begin{figure}[tbp]
\includegraphics[scale=0.48]{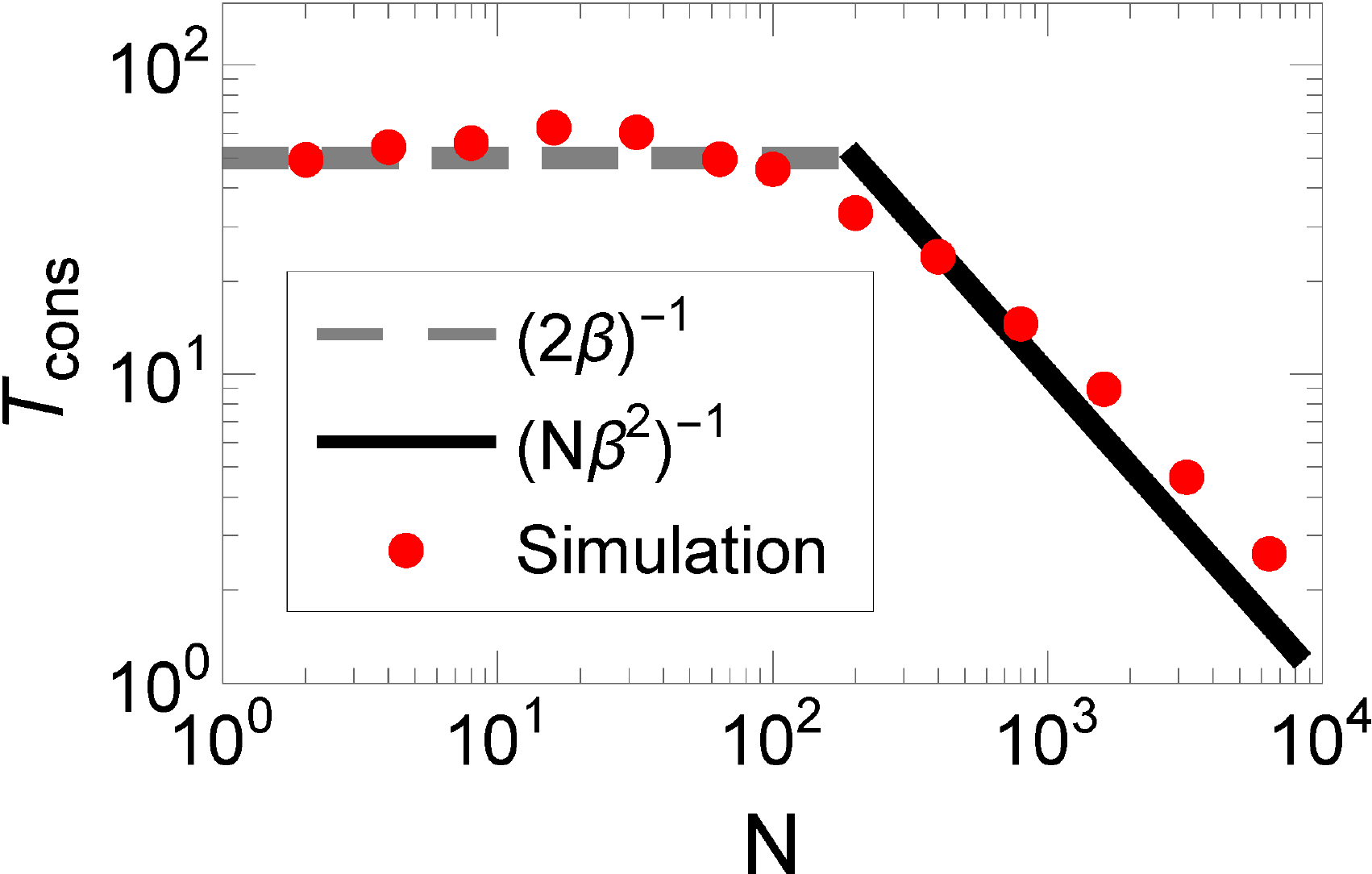}
\caption{\label{CompleteGraphConsensus}(Color online) Mean consensus time versus $N$ for a complete graph with $\beta = 10^{-2}$. Theory is the dashed gray line $T_\text{cons}\sim (2\beta)^{-1}$for small $N$, and the black line $T_\text{cons}\sim (N\beta^2)^{-1}$ for large $N$. This figure contrasts significantly with the IP model, which predicts that $T_\text{cons}\sim N$.}
\end{figure}

When $\delta = 0$, $\mu = 0$, and $\beta k = 1$, the CCIS model is similar to the invasion process (IP) \cite{VMConsensus}, in which a neighbor is randomly chosen to have the same opinion as the root node \cite{VMConsensus}. In the true IP, $T_\text{cons} \sim N \langle k^{-1} \rangle \langle k\rangle$, but in the CCIS model, $T_\text{cons} \sim \frac{N}{\langle k^2 \rangle}$ for large $N$. The discrepancy is due to a fixed fraction of neighbors, $1/\langle k\rangle$, being changed in the CCIS model, instead of exactly one in the IP. Interestingly, this implies that $T_\text{cons}\sim (N\beta^2)^{-1}$ in a complete graph, which we observe in Fig. \ref{CompleteGraphConsensus}, while in the IP, $T_\text{cons}\sim N$ for $N \ge 10$ (not shown). In the CCIS model, we find that, for small $N$, the consensus time is roughly $(2\beta)^{-1}$, the mean time for consensus to be reached between two nodes. The crossover to the asymptotic limit is when $T_\text{cons}=(2\beta)^{-1}= (N\beta^2)^{-1}$ or $N = 2/\beta$. In conclusion, although some of the scaling behavior resembles previous work on the VM, we make predictions that are completely distinct from previous VM-like models. 
This discrepancy has the potential to be tested in a social experiment by observing the time to consensus in small groups, because the difference is apparent even for small $N$. We leave this for future work.

\section{Consensus Times For $\delta > 0$}
\label{ConsensusTimes}

\begin{figure}[tbp]
\includegraphics[scale=0.58]{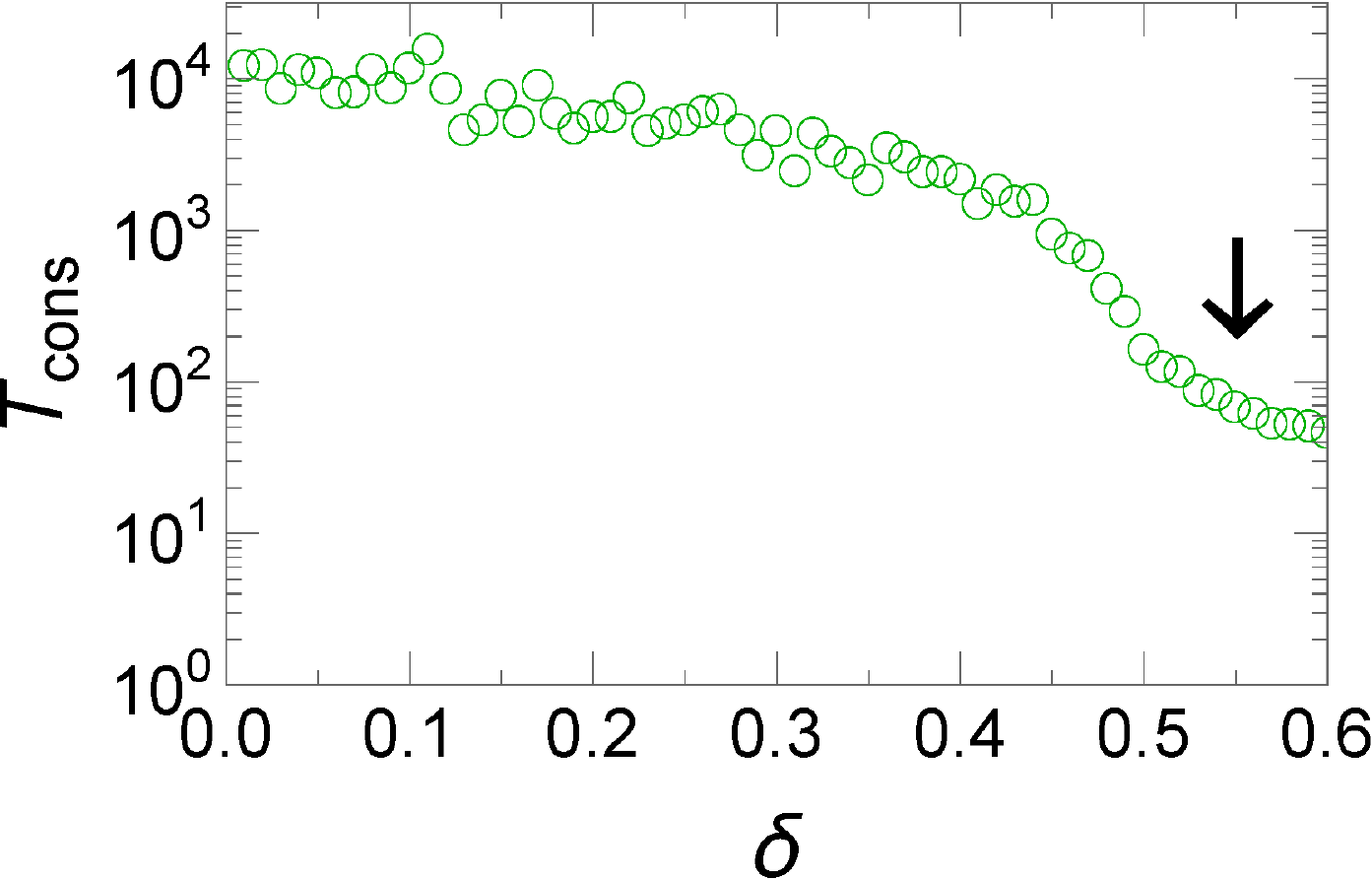}
\caption{\label{TvsDelta}(Color online) The consensus time versus $\delta$ with $\mu = 0$, and $\beta = 0.05$, on a $\langle k\rangle = 10$ Erdos-Renyi Network. The arrow indicates the critical point (calculated using SIS model analysis \cite{ScaleFreeSIS}) of the CCIS model, above which all individuals quickly approach the neutral state. We note that the consensus time appears to decrease monotonically with $\delta$. The initial condition is a 50/50 mixture of opinions 1 and 2.}
\end{figure}

\begin{figure}[tbp]
\includegraphics[scale=0.44]{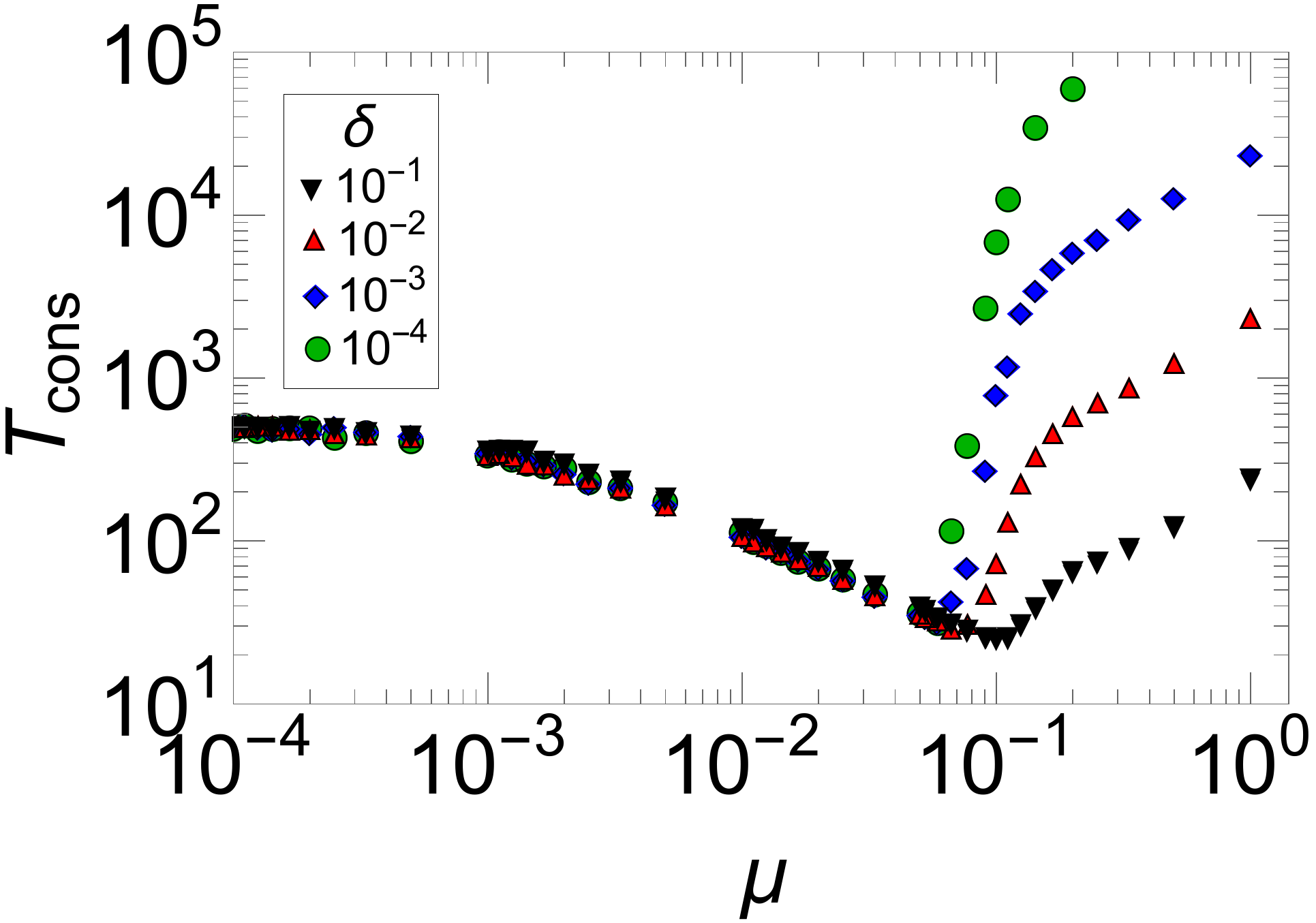}
\caption{\label{TtauMultiGamma}(Color online) Mean consensus time for varying $\mu$ and $\delta$ on $\langle k\rangle = 10,~N=10^4$ Poisson networks with $\beta = 0.5$. A minimum in the consensus time is observed for $\mu\approx 0.1$, while analysis of model behavior for $\mu > 0.1$ reveals that $T_\text{cons}\sim \text{log}(N) \delta^{-1}$.}
\end{figure}

Finally, we numerically study $T_\text{cons}$ for $\delta > 0$, where the previous analysis breaks down, in two ways. Figure \ref{TvsDelta} illustrates how the consensus time depends on the recovery rate $\delta$ when $\mu=0$. Figure \ref{TtauMultiGamma} shows how the consensus time depends on the stubbornness rate $\mu$ for different values of $\delta$.  Note that ``consensus'' here refers to the state in which at most one opinion remains. Thus the consensus state may contain a mixture of opinionated and neutral individuals, as long as all opinionated individuals hold the same opinion.

Fig. \ref{TvsDelta} shows that the consensus time decreases with $\delta$.Because the expected number of opinionated individuals at any given time decreases as $\delta$ increases, the time it takes for the opinionated individuals to reach consensus is also shorter.
For this reason, we hypothesize that $T_\text{cons}\sim N_\text{eff} \sum_A P^{(A)}$, with $N_\text{eff}$ as defined previously. In other words, we generalize Eq. \ref{DiffusiveConsensus} and claim $N_\text{eff} \sum_A P^{(A)}$ is the new effective size of the network which we leave for future work to explore more deeply.

In Fig. \ref{TtauMultiGamma}, we plot $T_\text{cons}$ versus $\mu$ for various values of $\delta$ to understand how our model more generally reaches consensus for finite networks. First, we find that $T_\text{cons}\sim \text{log}(N) \delta^{-1}$ for small $\delta$ and $\mu > 0.1$, which, in this limit, is in agreement with previous analysis \cite{FastConsensus}. The behavior of $T_\text{cons}$ versus $\mu$ demonstrates interesting parallels to other models \cite{FastConsensus,VMFreezePeriod,NoiseReducedVM} (Fig. \ref{TtauMultiGamma}), whereby at a non-trivial value of $\mu = \mu_c(\delta)$, the consensus time reaches a minimum, and at larger values of $\mu$ the consensus time increases significantly. This may generically imply that large groups reach consensus relatively quickly if individuals are moderately resistant to changing their opinion.

\section{Conclusion}
\label{Conclusion}

In conclusion, we have introduced a model of opinion dynamics that agrees with current empirical data and exhibits interaction dynamics based upon real human behavior. 

In addition, because our model makes few assumptions, it may plausibly explain a range of behaviors, which future empirical investigations may be able to corroborate. For example, the model can be used to explore the ``viral" spread of competing products, in which stubbornness is mapped to increasing brand loyalty\cite{BrandLoyalty1,BrandLoyalty2}. In this case, the brand-share distribution might be similar to Fig. \ref{QScaling}.

Future work is necessary, however, to model opinions with greater realism. 
As mentioned previously, this model might benefit from additional realistic assumptions. For example, mass media could be added, because it can be more influential than individual persons. Similarly, we could add party affiliation, which may bias which candidate(s) individuals initially prefer, or are likely to support in the future \cite{HeteroVM}. Additionally, the recovery rate could be tied to an individual's stubbornness, instead of constant as we assume here for simplicity. 

In addition, one could model heterogeneous stubbornness, either at the opinion level (as our model assumes) or individual level, because some individuals appear to stubbornly hold on to an idea, while others may shift their stance more readily. This is known to add greater realism to opinion dynamics because the most stubborn individuals possible, known as ``zealots" in previous literature, can help push the political preference in a two party system near the $50/50$ mark, alike to what we observe in the CCIS model \cite{Zealot1,Zealot2,Zealot3}. Expanding on previous work, we expect that adding heterogeneous stubbornness to our model can further slow down or stop consensus and potentially create better agreement with data. In addition, we assume agents linearly increase their resistance to alternative opinions in time. This is not necessarily true because PTP (pre-trial publicity) a day before a trial produced a negative correlation between the biased news and the juror decision, while PTP exactly a week before a trial is not statistically significant \cite{PTP}. A non-linear or non-monotonic stubbornness may significantly change the dynamics. 

Finally, this paper assumes that all opinions are equally strong and spread at the same time, but this is not necessarily true in reality, which we discuss briefly in Section \ref{ModelDetails}. MySpace started before Facebook, for example, and therefore more people initially preferred MySpace \cite{PathogenCompetitionWTA}. Facebook was later seen as a preferred option, however, and eventually dominated social media at the expense of MySpace and similar platforms. Future work should therefore allow for a first-mover advantage \cite{FirstMover} and opinions that are stronger or weaker than others to better capture reality.

\begin{acknowledgments}
We would like to thank Dr. Arnab Chatterjee for pointing us to the election distribution data.
Our work is supported by DARPA under contract No. N66001-12-1-4245 and No. D13PC00064.
\end{acknowledgments}

\appendix

\section{Fitting the CCIS Model to Data}
\label{fitting}

In this section we describe in more detail how the CCIS model is fit to empirical vote distribution data and correlation data

\subsection{Network Model}

To match the model to data, we use a spatially distributed network, which creates a non-zero spatial correlation, and we find that we need a scale-free distribution to best match scaled vote distribution data. Adding both of these properties to a single network, however, is not just convenient, but realistic. For example, we could try to run a model on the most natural spatial network: a grid. In a grid, individuals only interact if they are spatially close, but previous work on the ``six degrees of separation" between two randomly chosen individuals \cite{Milgram,6DegSeperation} and ``weak ties" between socially disparate individuals \cite{WeakTies}, suggests that ties can exist between individuals who are spatially separated by large distances. Furthermore, unlike grids, the degree distribution of many social networks is a power law \cite{BA}.

\begin{figure}[tbp]
\includegraphics[scale=0.43]{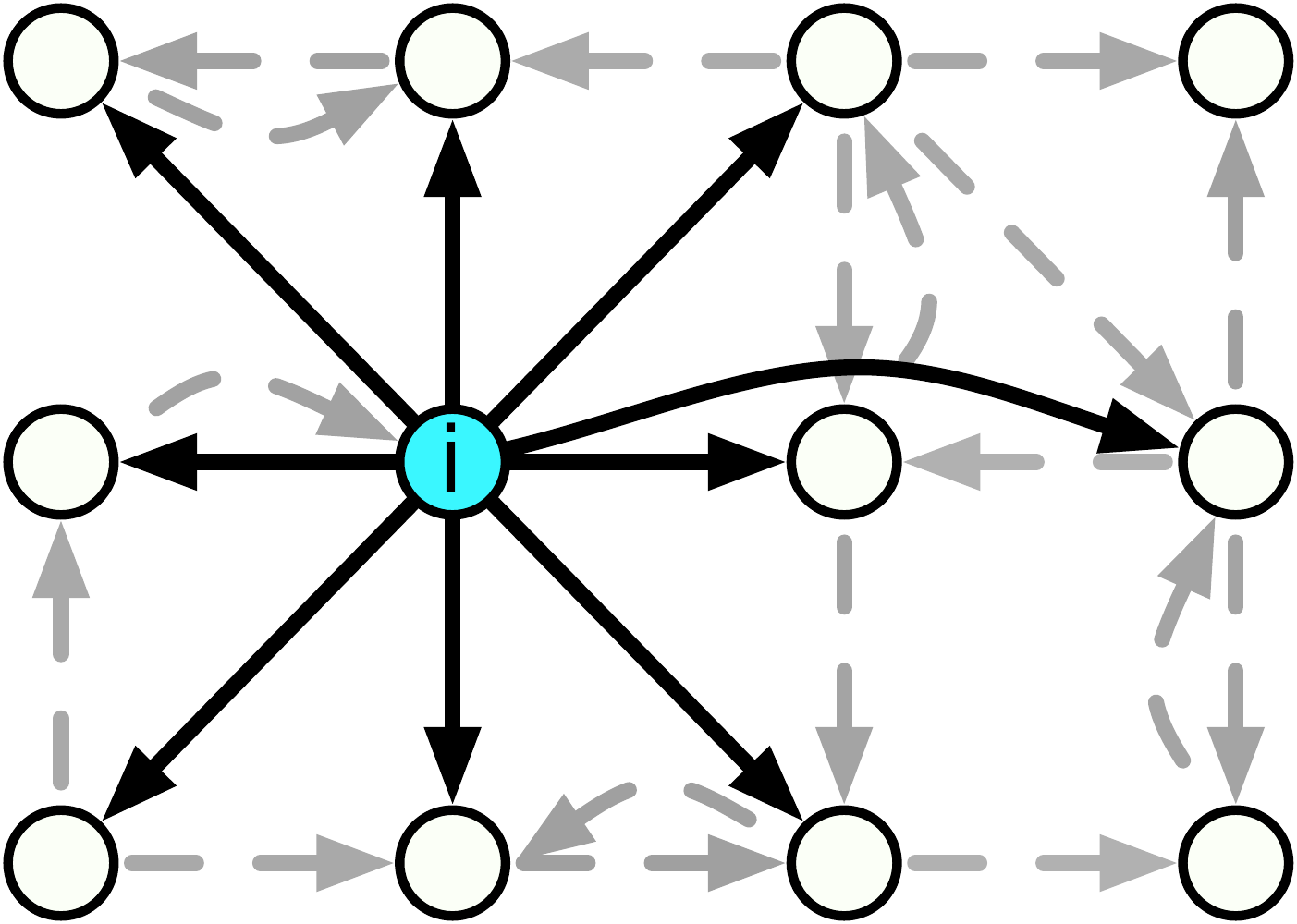}
\caption{\label{NetworkSchematic}(Color online) A schematic of the network chosen to fit our models to empirical data. All nodes have a scale-free out-degree distribution whereby a node $i$ with degree $k_i$ (in this example, $k_i = 9$) is then connected to its nearest neighbors.}
\end{figure}

Combining all these properties, we can create spatial scale-free networks, such as the one in Fig. \ref{NetworkSchematic}. Nodes have an out-degree $k$ chosen from a scale-free distribution, and are placed on a grid with unit distance. Each node is then connected to their nearest neighbors, although to test the robustness of our results, a fraction $f$ of 
 are randomly rewired. As $f$ increases, the model makes similar fits to the vote distribution data but the spatial correlation decreases. To keep $\langle k\rangle$ constant for fixed degree distribution $p(k > k_\text{min})$, we change the proportion of nodes with degree $k_\text{min}$ until we have the appropriate $\langle k\rangle$. 
The directed nature of the network reduces the chance of multi-edges or self loops, and it seems to be a reasonable assumption that people with a lot of connections broadcast their opinion to a wide audience 
without as much attention paid to the ideas of those same individuals.

\begin{figure}[tbp]

\includegraphics[width=0.46\textwidth]{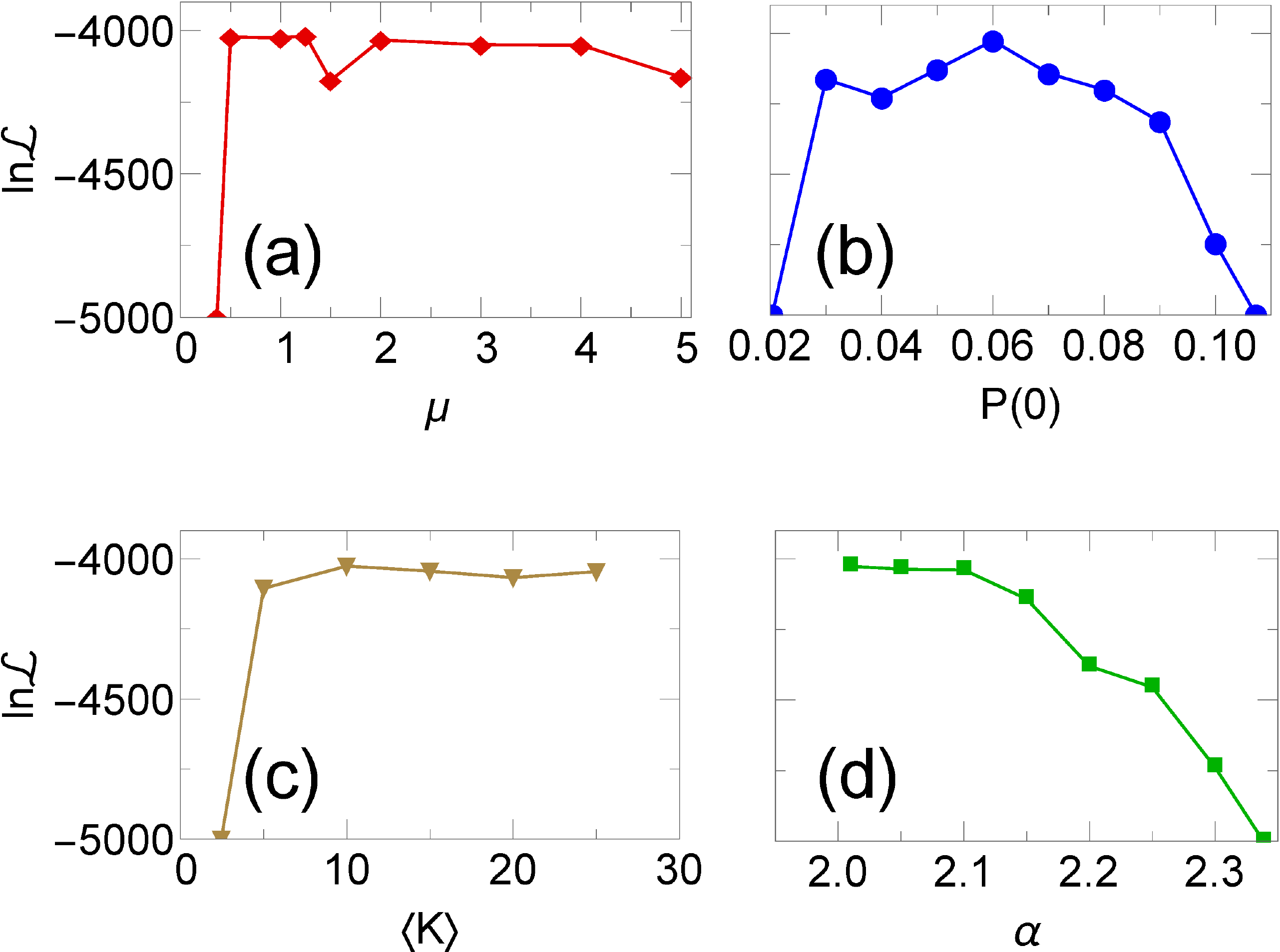}
\caption{\label{LE}(Color online) The log-likelihood function versus (a) $\mu$, (b) the fraction of individuals seeded, $P(0)$, (c) $\langle k\rangle$, and (d) $\alpha$. Not shown in (d) is the log-likelihood of a 10-regular spatial graph ($-10771$), which is far below the current y-axis scale. Arrows indicate the chosen values for our fit. $\mathcal{L}$ varying by less than $100$ does not look appear visually different from our fit. }
\end{figure}

\subsection{Fitting Model Parameters}
Next we discuss how our model is fit to data. The Poland 2005 data set is chosen due to the large number of elections (593, versus $\sim 200-400$ for other countries). In our simulations, seeded individuals are equally split among the various candidates, but variations in seeding should create similar results.  Maximum Likelihood Estimation (MLE) is used to determine the appropriate seeding fraction.



The model has no readily apparent closed-form solution, and a Kernel Density Estimator for the model greatly over-estimates the probability for small $x_i$, therefore we approximate the probabilities with log-binned histograms (the widths, however, do not seem to change the best fit parameter significantly).

\subsection{Parameter Values}

In the FC model, only the candidate has an initial preference of whom to vote for, while in our model, we assume a set percentage of individuals have an initial preference to some candidate. The CCIS model creates a poorer fit when $Q$ individuals are seeded (not shown), but seeding a fixed percentage seems to be an equally realistic assumption if we imagine that a small percentage of voters are initial strong supporters of the candidates.

We can also let the fraction seeded be 100\%. Holding $\mu = 1$, the best fit $\beta$ value is $0.65$, with a fit similar to Fig. \ref{QScaling} (see Fig. \ref{InitInfComparison}). We choose to seed less than the total population, because it seems reasonable that at some starting point, not everyone is aware of the candidates.

To fit our model to the distributions, we set $\beta$ to $0.1$, $\mu$ to $1$, and $\langle k\rangle$ to 10, but variations in these values do not significantly affect our results (see Fig. \ref{LE} in Appendix \ref{fitting}, where we hold all parameters fixed, except for the given parameter plotted). We also fix $\delta = 0$ in order for the distribution to remain fixed in equilibrium.
The MLE for alpha, however, varies depending on the type of network chosen. For example, while $\alpha  = 2.01$ creates a good fit with our current model network (spatially-extend scale-free),  $\alpha = 2.5$ creates a good fit on an undirected scale-free network with no spatial structure.
Whatever the optimal $\alpha$, however, we find that a wide distribution (e.g., $\alpha < 3$) works best, when fitting to data. A Poisson or $k$-regular graph, for example, never appears to fit well with data, regardless of the other parameter choices.

We have more freedom to vary all parameters if our only goal is to create vote correlations similar to empirical data (Fig. \ref{SpatialCorrel}). The roughly logarithmically decreasing correlation with distance is observed for many values of $\delta \ge 0$, $\mu\ge0$, $\beta > 0$, $P(0)>0$ and $\alpha > 2$. Just one example are the parameters chosen in Fig. \ref{SpatialCorrel}.

\subsection{Determining The Spatial Correlation}

We finally mention how the correlations in Fig. \ref{SpatialCorrel} are calculated. To be consistent with previous work \cite{CorrelationData,CorrelationData2} and Fig. \ref{SpatialCorrel}a, we define the normalized correlation in our figures as:
\begin{equation}
C(r) = \frac{\langle P^{(1)}_i P^{(1)}_{j} | d_{ij} \approx r\rangle-\langle P^{(1)}_i \rangle^2}{\sigma_{P^{(1)}}^2},
\label{RegionCorrel}
\end{equation}
in which $P^{(1)}_i$  is the fraction of voters for candidate $1$ within a small region (which we choose to be $5\times5$ node squares), $\langle P^{(1)} \rangle$ is the average fraction of voters for candidate $1$, and $\sigma_{P^{(1)}}^2$ is the variance in vote distribution across all regions. $\langle P^{(1)}_i P^{(1)}_{j} | d_{ij} \approx r\rangle$ is the 2-point correlation function between regions whose centroid is a distance $r\pm 1/2$ from each other.

\section{Derivation of the Transport-Like Approximation}
\label{TLA}

In this appendix, we use the TLA to understand the initial jump in the opinion densities (see Figs. 6 \& 12). 

\begin{figure}[tbp]
\includegraphics[scale=0.43]{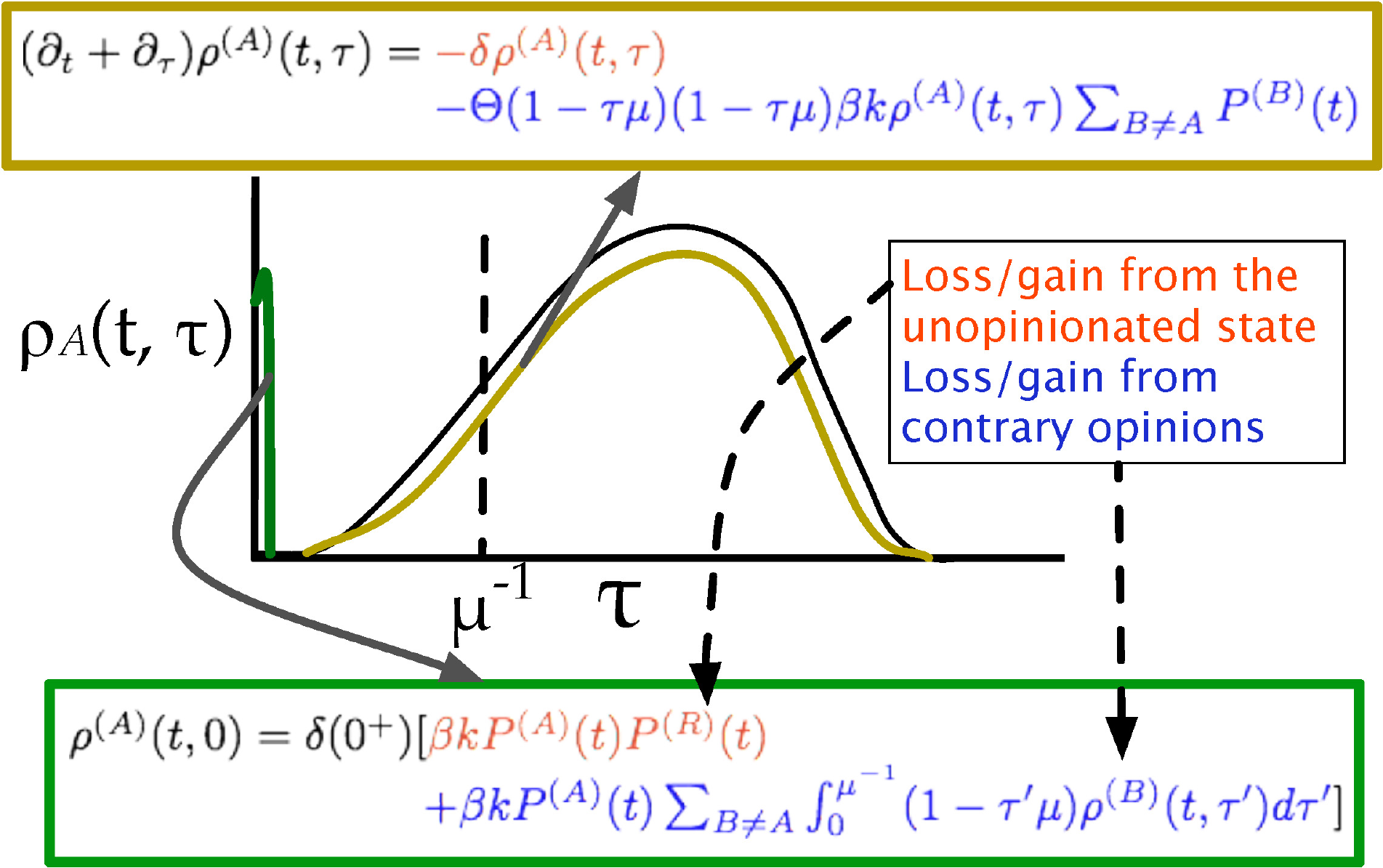}
\caption{\label{MFTSchematic}(Color online) Schematic of the scalar variable in Eq. \ref{MFTNew} as a function of time, $t$, and time since opinion adoption, $\tau$.}
\end{figure}

Our model can be described by the following equation in the mean field:


\begin{widetext}
\begin{equation}
\label{MFTNew}
(\partial_t + \partial_{\tau} ) \rho^{(A)}(t,\tau) = {\color{LightRed}-\delta \rho^{(A)}(t,\tau)} 
{\color{blue}- \Theta(1-\tau \mu) (1 - \tau\mu)\beta k\rho^{(A)}(t,\tau)\sum_{B\neq A} P^{(B)}(t)},
\end{equation}
with the boundary conditions:
\begin{enumerate}
\item $\rho^{(A)}(t,\infty) = 0$,
\item$\rho^{(A)}(t,0) = \delta (0^+)[
\textcolor{LightRed}{\beta k P^{(A)}(t) \tilde{P}(t)} 
{\color{blue}
+\beta k P^{(A)}(t)\sum_{B\neq A}\int_0^{\mu^{-1}} (1-\tau' \mu)\rho^{(B)}(t,\tau') d\tau'
}
]$,
\item and $\rho^{(A)}(0,\tau) = f(\tau)$,
\end{enumerate}
\end{widetext}
(see Fig. \ref{MFTSchematic} for a visual representation) where $t$ is time, $\tau$ is the time an individual has had their most recent opinion, $P^{(X)}$ is the fraction of individuals with opinion $X$ at time $t$, $\rho^{(X)}$ is the density of individuals with opinion $X$  at time $t$ who have kept their opinion for a time $\tau$ (variables and parameters are also defined in Table I). Finally, $\tilde{P}(t)  = 1-  \sum_{X}P^{(X)}(t)$. The RHS of the equation describes the ability of individuals to recover (light grey or \textcolor{LightRed}{red} term) as well as the ability to change opinions (dark gray or {\color{blue}blue} term). We can interpret the boundary conditions as:
\begin{enumerate}
\item Normalizability
\item An increase in the infection density due to neutral neighbors (light gray or \textcolor{LightRed}{red} term) and 
opinionated neighbors (dark gray or {\color{blue}blue} term).
\item Initial conditions
\end{enumerate}

 We focus on the simpler case of $\delta = 0$ for our analysis because adding $\delta$ to the equation numerically does not seem to affect consensus, while, in simulations, consensus happens quickly. Future analysis of perturbations around equilibrium, however, may give us better insight into what happens in simulations. We do know, however, that when $\delta \ll 1$, the equation can be simplified to the one seen in \cite{FastConsensus}, were they find, to use our notation, $T_\text{cons}\sim\delta^{-1}$, in agreement with our own simulations (not shown).

The simplified equation is:
\begin{widetext}
\begin{equation}\tag{2}
(\partial_t + \partial_{\tau} ) \rho^{(A)}(t,\tau) =
- \Theta(1-\tau \mu) (1 - \tau\mu)\beta k\rho^{(A)}(t,\tau)\sum_{B\neq A} P^{(B)}(t)
,
\end{equation}
\end{widetext}

\noindent
with the same boundary conditions.

We first try to understand the transient ``jump" in the fraction of individuals following a given opinion on a timescale that is in many cases much smaller than the time to reach consensus. We wish to understand the the equilibrium fraction of individuals with a given opinion, and the time to reach equilibrium.

We find strong agreement between theory and simulations for the equilibrium fraction of individuals in each opinion (Fig. \ref{TheoryVsSim}), especially when $\langle k\rangle \ge 10^2$. For fixed networks with $\langle k\rangle <10^2$, the equilibrium values are on average below theoretical values, plausibly because individuals are less connected to their neighbors, and thus less influenced by them, than the mean field theory assumes. To find agreement with simulations, we numerically determined equilibrium values by stepping forward the equation using the forward Euler method.

\begin{figure}[tbp]
\includegraphics[scale=0.55]{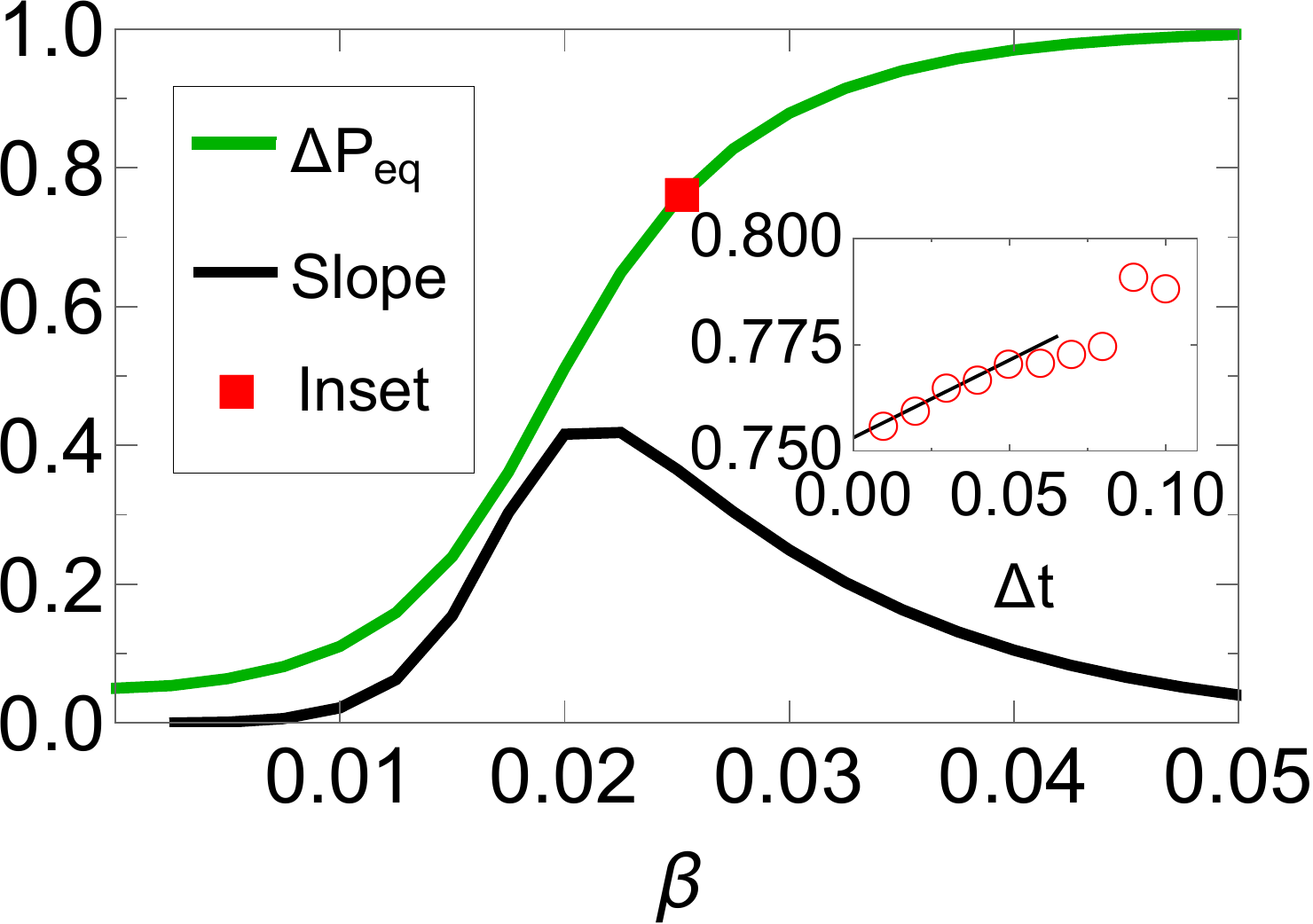}
\caption{\label{NumericalCalc}(Color online) Details regarding the theory curve of Figure \ref{TheoryVsSim}. (Inset) For each value of $\beta$, we vary the timestep width for Eq. \ref{MFT} ($\Delta t$), and find the resulting equilibrium value. $\Delta P_\text{eq}(\Delta t\rightarrow 0)$ is estimated via linear regression. (Main figure) Plotting $\Delta P_\text{eq}(\Delta t\rightarrow 0)$ and slope for $\Delta P(0) = 0.05$, we find the slope, seen in the inset, is greatest when $\Delta P_\text{eq} \approx 0.6$, implying the error from using a single value of $\Delta t$ would have been largest in this range.}
\end{figure}

This method is inherently sensitive to the timestep width, $\Delta t$, especially when $\Delta P_\text{eq} \approx 0.5$, therefore we find the equilibrium value can be more accurately determined by varying the timestep width and, via linear regression, determining the asymptotic limit for the equilibrium as $\Delta t\rightarrow 0$ (Fig. \ref{NumericalCalc}).
 This seems to reduce our statistical error to less than $0.5\%$ compared to as much as $1-6\%$, and is in excellent agreement with the simulations.

Next, we determine the time to reach equilibrium. We discretize $\tau$, following \cite{FastConsensus}, to derive a set of equations that we linearize around a fixed point to determine the scaling of the transient time (Eq. \ref{eval} \& \ref{Teq}). Our approximations are only accurate for $\mu\ll1$, but seem to be qualitatively similar to numerical data for $\mu\sim O(1)$. We define the following macroscopic variables:

\begin{equation}
P^{(1)}(t) = \sum_{\tau'} \rho^{(1)}(t,\tau'),\text{ and } P^{(2)}(t) = \sum_{\tau'} \rho^{(2)}(t,\tau'),
\end{equation}
in which $\sum_{\tau'}$ is shorthand for $\sum_{\tau' = 0}^\infty$.
 If we let $\Omega[|]$ be the conditional probability function, and $\dot x\equiv \frac{d}{dt} x$, then Eq. \ref{MFT} becomes (for $\tau > 0$):
\begin{equation}
\begin{split}
\dot\rho^{(1)}(t,\tau)  ~~~~~~~~~~~~~~~~~~~~~~~~~~~~~~~~~~~~~~~~~~~~~~~~~~~~~~~~~~~~~\\
= \Omega[\rho^{(1)}(t,\tau) | \rho^{(1)}(t,\tau-1)] \rho^{(1)}(t,\tau-1) - \rho^{(1)}(t,\tau) ~~~~~
\end{split}
\end{equation}

Expanding these variables out, we find that:

\begin{widetext}
\begin{equation}
\dot\rho^{(1)}(t,\tau)=\text{{\bf(}}1+ \beta k \{[P^{(1)}(t) + \mu (\tau - 1)] P^{(2)}(t)-1\}\text{{\bf)}} \rho^{(1)}(t,\tau - 1) - \rho^{(1)}(t,\tau),
\end{equation}
\end{widetext}
and for $\tau = 0$:
\begin{equation}
\dot \rho^{(1)}(t,0) = \beta k P^{(1)}(t) [P^{(2)}(t) - I^{(2)}(t)] - \rho^{(1)}(t,0).
\end{equation}
With an equivalent set of equations for $\rho^{(2)}(t,\tau)$ and
\begin{equation}
I^{(1)}(t) = \sum_{\tau'} \mu \tau' \rho^{(1)}(t,\tau'), ~I^{(2)}(t) = \sum_{\tau'} \mu \tau' \rho^{(2)}(t,\tau').
\end{equation}
From the above results we can sum $\rho^{(1)}(t,\tau)$ to find the equations for the macroscopic variables:
\begin{equation}
\dot P^{(1)}(t) = \beta k [I^{(1)}(t) P^{(2)}(t) - I^{(2)}(t) P^{(1)}(t)].
\end{equation}

To lowest order in $\mu$, we also find that:
\begin{equation}
\begin{split}
\dot I^{(1)}(t) \approx \mu (1-\beta k) P^{(1)}(t)~~~~~~~~~~~~~~~~~~~~~~~~~~~~~\\
 + \beta k [\mu P^{(1)}(t)^2 + I^{(1)}(t) P^{(1)}(t) - I^{(1)}(t)].
\end{split}
\end{equation}
These equations are solvable by expanding around the solution $P^{(1)}= P^{(2)}  = 1/2$ and $I^{(1)} = I^{(2)} = \mu[1/(\beta k) -1/2]$ to first order. The resulting largest eigenvalue is 

\begin{equation}
\begin{split}
\lambda_1 \approx~~~~~~~~~~~~~~~~~~~~~~~~~~~~~~~~~~~~~~~~~~~~~~~~~~~~~~~~~~~~~~~~~~~\\
 \frac{-\beta k - 4 \mu + 2 \beta k \mu + \sqrt{16 \beta k \mu + (\beta k + 4 \mu - 2 \beta k \mu)^2})}{4} 
\label{eval}.
\end{split}
\end{equation}

\begin{figure}[tbp]
  \includegraphics[scale=0.43]{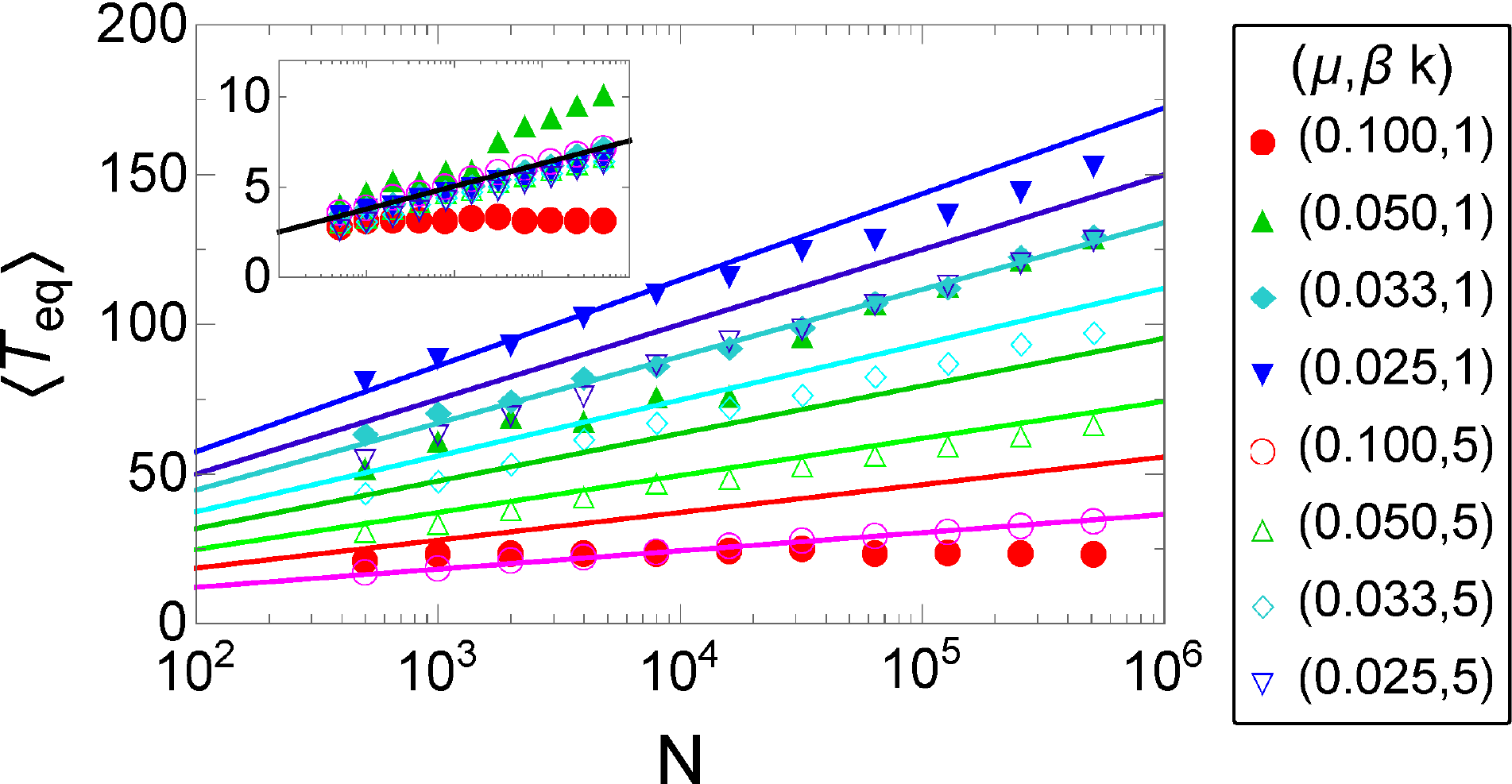}
\caption{\label{TeqPlot}(Color online) $T_\text{eq}$ versus $N$ for various $\mu$ and $\beta k$ (simulations on $k$-regular random graphs,with $k = 10$). Inset shows collapse when $T_\text{eq}$ is rescaled by $\lambda_1$, with the best fit slope equal to $\nu$ in Eq. \ref{Teq}.}
\end{figure}

 The time to reach equilibrium, $T_\text{eq}$, is: 
\begin{equation}
T_\text{eq} = \nu \frac{\text{log}(N)}{\lambda_1}
\label{Teq},
\end{equation}

\noindent
where $\nu$ is a fitting parameter found to be $1.26\pm 0.04$ from simulations. 
When $\beta k = 1$, this eigenvalue should agree exactly with the value cited previously \cite{FastConsensus}, but we find disagreement by an overall prefactor of $1/4$ which, at least to our knowledge, may have been missed in the previous work. Figure \ref{TeqPlot} shows how simulations agree with theory. We notice disagreement is most significant when $\mu$ approaches 1, and $\beta$ is small (e.g., $\beta = 0.1$).

\section{Scaling of effective network size}
\label{FPA}

In this appendix we derive Eq. \ref{DiffusiveConsensus} using a FPA, which is distinct from the TLA in Appendix \ref{TLA}. Our derivation is heavily based on the derivation of consensus times for the VM and Invasive Process by Sood, Antal, and Redner \cite{VMConsensus}. 

\subsection{Derivation}

We use the same conventions as in that paper, except the transition probability scaling factor, $S$, the degree distribution, $p(k)$, and associated moments, $\langle k^m \rangle = \sum_k p(k) k^m$.  Note that we assume for now that $\beta < 1/\langle k \rangle$ for the inward spreading process (opinions spread inward toward an individual).

Let $\eta(x)$ be the state of a node $x$ on a network with adjacency matrix $A_{xy}$ and order $N$. Assuming 2 opinions and that $\mu$ and $\delta \rightarrow 0$, we have a two-state system. Using the conventions of Sood, Antal, and Redner \cite{VMConsensus} the opinions of the two-state system are ``0" or ``1" instead of ``1" or ``2". We stress that the $0$ state {\em is} an opinionated state. Lastly $\eta_x$ is the state of the system after changing a node $x$:
\begin{equation}
\eta_x(y) = \begin{cases}
\eta(y), & y \neq x \\
1-\eta(x), & y = x
\end{cases}.
\end{equation}
The transition probability at node $x$ is therefore:
\begin{equation}
\text{{\bf P}}(\eta \rightarrow \eta_x) = \sum_y \frac{A_{xy}}{N S} [\Phi(x,y) + \Phi(y,x)],
\end{equation}
in which:
\begin{equation}
S = \beta^{-1},
\end{equation}
and
\begin{equation}
\Phi(x,y) = \eta(x) [1-\eta(y)].
\end{equation}

We further assume a mean field solution, in which the adjacency matrix becomes the average adjacency matrix:
\begin{equation}
A_{xy} \rightarrow \langle A_{xy}\rangle = \frac{k_x k_y}{\langle k \rangle N}.
\end{equation}
Instead of individual states $\eta(x)$, we can instead focus on $\rho_k$, the density of states with degree $k$:
\begin{equation}
\rho_k = \frac{1}{N} \sum_{x'} \eta(x').
\end{equation}
Here, $x'$ is the sum of all nodes with degree $k$. To clarify the below equations, we  also define a variable $\omega$:
\begin{equation}
\omega = \frac{1}{N \langle k \rangle} \sum_x k_x \eta (x) = \frac{1}{\langle k \rangle} \sum_k k p(k) \rho_k.
\end{equation}

Next, we define our raising and lowering operators for $\rho_k$, which defines the probability of increasing or decreasing $\rho_k$ by a small increment:
\begin{equation}
\rho_k \rightarrow \rho_k^{\pm}  \equiv \rho_k \pm \delta\rho_k,
\end{equation}
 in which 
\begin{equation}
\delta\rho_k = 
\begin{cases}
\frac{N \langle k^2\rangle }{S \langle k\rangle p(k)} & \text{Outward Process}\\
\frac{N \langle k\rangle }{S p(k)} & \text{Neutral Process}\\
\frac{N k }{S p(k)} & \text{Inward Process}\\
\end{cases}
\end{equation}
The change in $\rho_k$ is proportional to the probability that an individual of degree $k$ changes his or her opinion in a given time step. This probability scales as $\frac{\langle k^2\rangle}{\langle k\rangle}$, $\langle k\rangle$, and $k$ for the outward, neutral, and inward processes, respectively.

The raising operator is defined as:
\begin{equation}
\textbf{R}_k=\text{{\bf P}}(\rho_k \rightarrow \rho_k^+) = \sum_{x'} \sum_y \frac{k_{x'} k_y}{S \langle k \rangle N^2} \Phi (y,x).
\end{equation}
With simplification, this yields
\begin{equation}
\text{\bf R}_k = \frac{\omega}{S} p(k) k (1 - \rho_k).
\end{equation}
Similarly, for the lowering operator:
\begin{equation}
\text{\bf L}_k=\text{\bf P}(\rho_k \rightarrow \rho_k^-) = \frac{\rho_k}{S} p(k) k (1 - \omega).
\end{equation}
The exit probability, $\xi_1$, defined as the probability for all nodes to reach state one in equilibrium, is the same for all cases
\begin{equation}
\xi_1 =\langle\rho\rangle \equiv\sum_k \rho_k p(k),
\end{equation}
and similarly, that $\langle\rho\rangle$ (or magnetization, if this were a spin system) is a conserved quantity. The reason is because
\begin{equation}
\begin{split}
\langle \Delta \eta(x)\rangle = [1-2 \eta(x)] \text{{\bf P}}(\eta \rightarrow \eta_x) \\
=  [1-2 \eta(x)]\sum_y \frac{A_{xy}}{N S} [\Phi(x,y) + \Phi(y,x)],
\end{split}
\end{equation}
\begin{equation}
\Delta\langle \rho\rangle = \sum_x \langle \eta(x)\rangle = \sim \sum_{x,y}[\eta(x) - \eta(y)],
\end{equation}
which is trivially 0. We note that this argument is exact (not a mean field approximation) and is independent of the method in which opinions spread (at least, again, assuming $\beta < 1/\langle k\rangle$ for the inward dynamics). 
The time to consensus is
\begin{equation}
\begin{split}
T_\text{cons}(\{\rho_k\}) =\sum_k  \Delta t_k~~~~~~~~~~~~~~~~~~~~~~~~~~~~~~~~~~~~ \\
+ [\text{\bf R}_k(\{\rho_k\}) T_\text{cons}(\rho_k^+) +\text{\bf L}_k(\{\rho_k\}) T_\text{cons}(\rho_k^-)]\\
+[1 - \sum_k \text{\bf R}_k(\{\rho_k\})+\text{\bf L}_k(\{\rho_k\})] T_\text{cons}(\{\rho_k\})
\end{split}.
\end{equation}
The average number of interactions per timestep is:
\begin{equation}
\Delta t_k = 
\begin{cases}
p(k) \frac{\langle k^2 \rangle}{\langle k\rangle S N} & \text{Outward Process} \\
p(k) \frac{\langle k \rangle}{S N} & \text{Neutral Process} \\
\frac{k }{S N} & \text{Inward Process} \\
\end{cases}.
\end{equation}
We expand to second order in $\Delta \rho_k$ and find that
\begin{equation}
\sum_k v_k \frac{\partial T_\text{cons}}{\partial \rho_k} + D_k\frac{\partial^2 T_\text{cons}}{\partial \rho_k^2} = -1,
\end{equation}

\begin{figure}[tbp]
\includegraphics[scale=0.58]{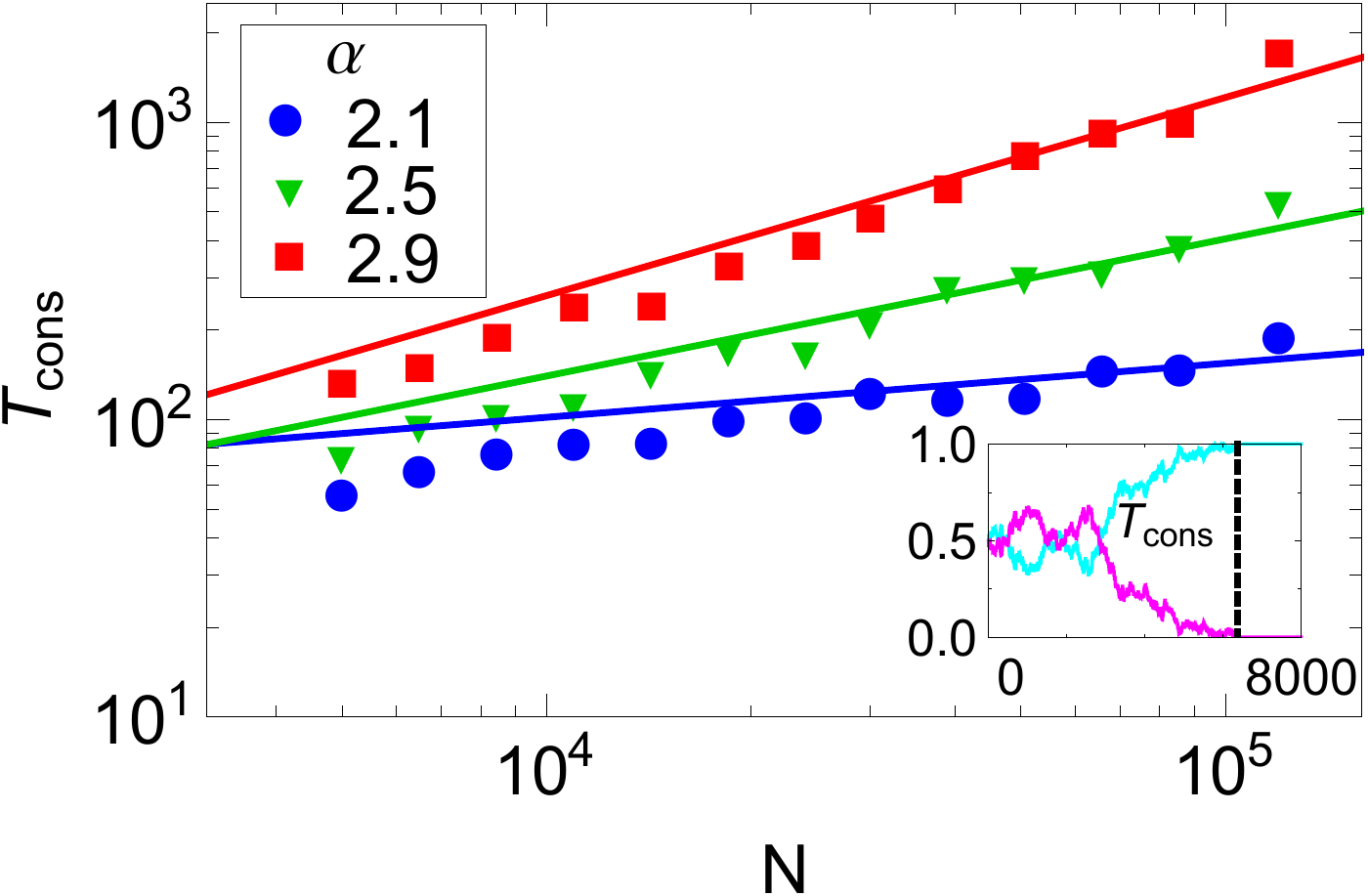}
\caption{\label{Tn}(Color online) Mean consensus time, $T_\text{cons}$, for scale-free networks with $\beta = 0.5$, $\delta = 0$, and $\mu = 0$. Inset is one example of consensus with $P^{(1)}(t)$ and $P^{(2)}(t)$. Using the FPA, the expected fit (solid lines) is Eq. \ref{DiffusiveConsensus}. Simulations are averaged over 10 networks (30 networks for $3 \times 10^4\le N < 10^5$, and 20 networks for $N= 1.2\times 10^5$) with 100 trials per network.}
\end{figure}

\noindent
in which
\begin{equation}
v_k \equiv \frac{\Delta \rho_k}{\langle \Delta t \rangle} (\text{\bf R}_k - \text{\bf L}_k) \rightarrow 0.
\end{equation}
As is shown in in the original voting model paper \cite{VMConsensus}, this value reaches 0 for time $T_\text{cons}\sim O(1)$ which is much smaller than the next term:
\begin{equation}
D_k \equiv \frac{(\Delta \rho_k)^2}{\langle \Delta t \rangle} \frac{(\text{\bf R}_k + \text{\bf L}_k)}{2}.
\end{equation}

A change of variables implies that:
\begin{equation}
\frac{\partial T_\text{cons}}{\partial \rho_k}  = \frac{\partial T_\text{cons}}{\partial \rho}  \frac{\partial \rho}{\partial \rho_k} = p(k) \frac{\partial T_\text{cons}}{\partial \rho},
\end{equation}
therefore
\begin{equation}
\sum_k \frac{M}{2 \langle k \rangle N S^2} (\omega + \rho_k - 2 \omega \rho_k) p(k)\frac{\partial^2 T}{\partial \rho^2}  = -1.
\end{equation}
in which
\begin{equation}
M = 
\begin{cases}
\langle k^2 \rangle & \text{Outward Process} \\
\langle k \rangle^2 & \text{Neutral Process} \\
k^2& \text{Inward Process} \\
\end{cases}.
\end{equation}
This can be made into a more compact form, noting that $\rho$ is conserved and $v_k \rightarrow 0$, $\omega \rightarrow \rho$. :
\begin{equation*}
\frac{\rho (1-\rho)}{N_\text{eff}}\frac{\partial^2 T_\text{cons}}{\partial \rho^2}  = -1.
\end{equation*}
[Eq. \ref{FPEqu} from the main text], where $N_\text{eff}$ is as follows:
\begin{equation*}
N_\text{eff} = \frac{N S^2}{\langle M \rangle} = 
\begin{cases}
\frac{N}{\beta^2 \langle k^2 \rangle} & \text{Outward Process} \\
\frac{N}{\beta^2 \langle k \rangle^2} & \text{Neutral Dynamics} \\
\frac{N}{\beta^2 \langle k^2 \rangle} & \text{Inward Dynamics} \\
\end{cases}.
\end{equation*}
\noindent
[Eq. \ref{DiffusiveConsensus} from the main text].

We find that this equation simplifies down to (24) in \cite{VMConsensus}, noting the boundary condition, $T_\text{cons}(0) = T_\text{cons}(1) = 0$ in which:
\begin{equation}
T_\text{cons}(\rho) = N_\text{eff} \left[(1-\rho) \text{ln}\frac{1}{1-\rho} + \rho \text{ln}\frac{1}{\rho}\right],
\end{equation}
implying that $T_\text{cons} \sim N_\text{eff}$.

As we discuss shortly, if $\beta > 1/\langle k\rangle$ in the inward-spreading case, we have VM dynamics, and the mean field consensus time replaces $\beta$ with $1/\langle k\rangle$. Furthermore, this approximation breaks down for small $\langle k\rangle$ and small $\beta$, in which we show in Section \ref{Analysis} that the consensus time scales as $\beta^{-1}$.
Future work could improve the accuracy of the current results with a pair approximation theory \cite{PAHomo,PAHetero}.

\begin{figure}[tbp]
\includegraphics[scale=0.5]{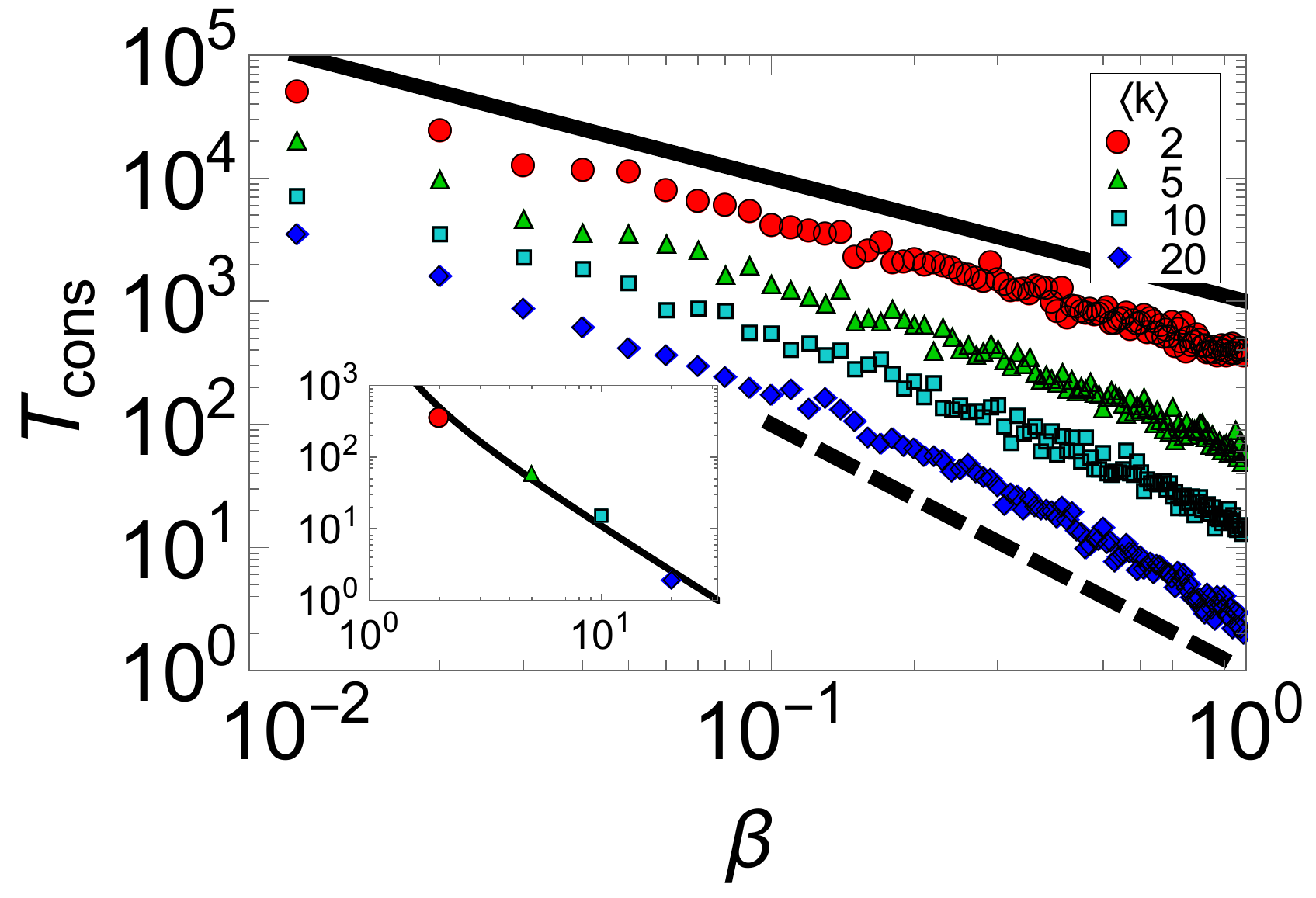}
\caption{\label{Tbeta}(Color online) Mean consensus time for varying $\beta$ ($\delta$ and $\mu = 0$) on 1000 node Poisson networks with different average degree, $\langle k\rangle$. Inset: consensus time versus average degree for $\beta = 0.99$. Simulations are averaged over 30 networks. The theory are the dashed lines ($T_\text{cons} \sim \beta^{-2}$ when $\beta$ and $\langle k\rangle$ large, and $T_\text{cons} \sim \beta^{-1}$ in the opposite limit) and solid line in the inset ($T_\text{cons}\sim \langle k^2\rangle^{-1}$).}
\end{figure}

\begin{figure}[tbp]
\includegraphics[scale=0.6]{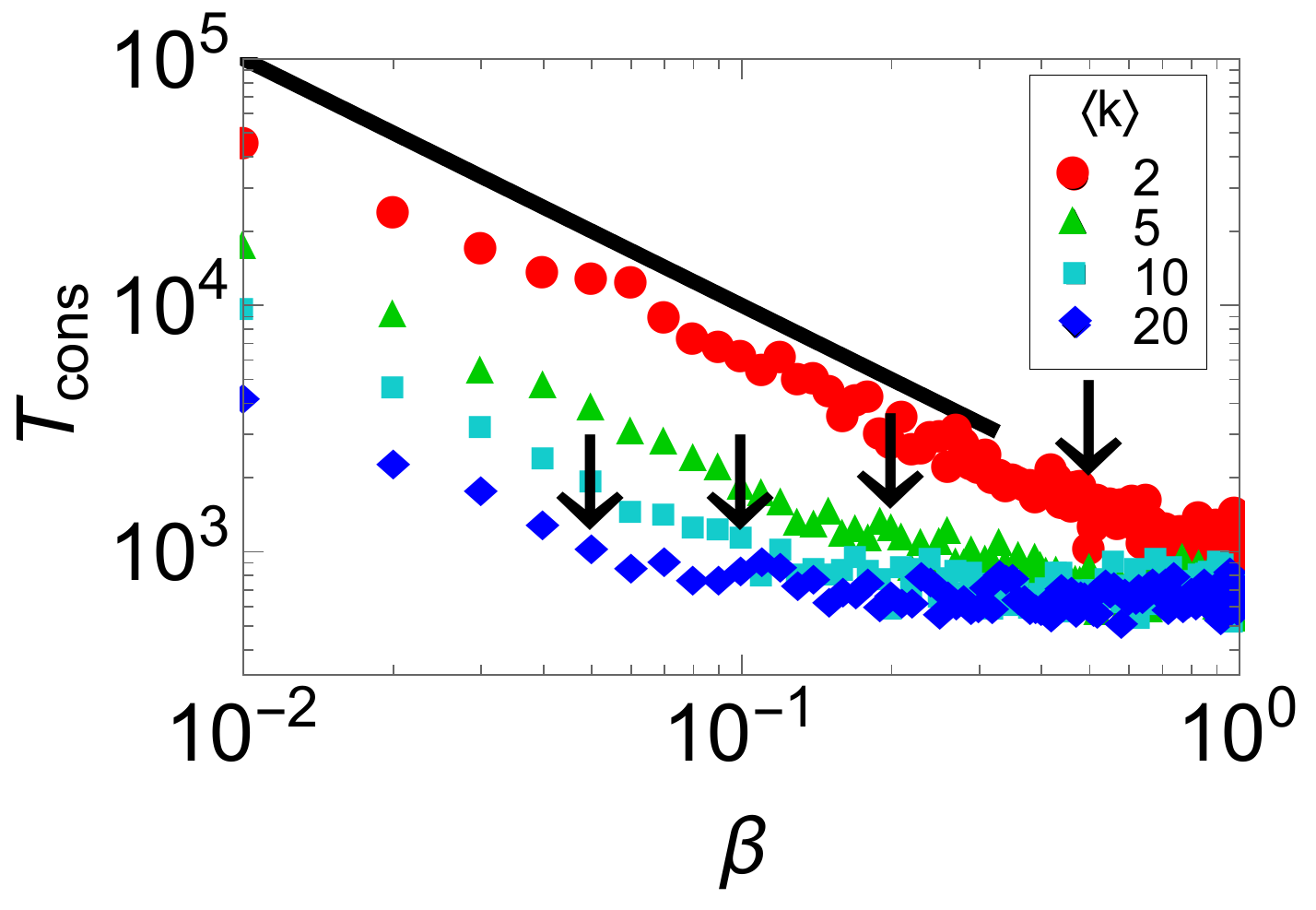}
\caption{\label{VMDynamicsConsensus}(Color online) Mean consensus time versus $\beta$ for Poisson graphs in which we use the inward infection method. Theory is the black line $T_\text{cons}\sim \beta^{-1}$ and arrows indicate when $\beta\langle k\rangle = 1$, whereby we transition to true VM dynamics, which is independent of $\beta$.}
\end{figure}

This paper mainly focuses on the outward process, but we have also compared theory and simulation for the other processes by varying $\beta$ and $\langle k^2\rangle$, in Poisson and scale-free networks, while setting $\delta$ and $\mu$ to $0$. First, we observe the dependence on $\langle k^2\rangle$ by simulating the models on scale-free networks.

In a scale-free network, $\langle k^2\rangle$ diverges with network order, $N$:
\begin{equation}
 \langle k^2 \rangle\sim
\begin{cases}
N^{3-\alpha} & \alpha < 3\\
\text{log}(N) & \alpha = 3 \\
O(1) & \alpha > 3 
\end{cases}.
\end{equation}
Therefore, for outward and inward dynamics:
\begin{equation}
T_\text{cons}\sim
\begin{cases}
O(1) & \alpha < 2 \\
\text{log}(N)^2 & \alpha = 2 \\
N^{2 (\alpha-2)/(\alpha - 1)} & 2 < \alpha < 3\\
N/\text{log}(N) & \alpha = 3 \\
N & \alpha > 3 
\end{cases}.
\end{equation}

\subsection{Agreement With Simulations}

Fig. \ref{Tn} compares outward process simulations to the FPA (inward processes simulations are similar, due to the equilvalent scaling). Although a finite size transient impedes this scaling behavior for $N \le 10^4$, we still see agreement for large enough networks. For Poisson networks, we see $T_\text{cons}\sim \langle k^2\rangle^{-1} = (\langle k\rangle^2 - \langle k \rangle)^{-1}$ in the inset of Fig. \ref{Tbeta}.

The inward-spreading dynamics closely parallel the outward spreading dynamics when $\beta^2\langle k^2\rangle < 1$. On the other hand, when $\beta$ is large enough, each node is, on average, infected by multiple nodes at each timestep, although, by the end of the timestep, only one opinion is chosen. This maps exactly onto the VM, and therefore so does the consensus time (Fig. \ref{VMDynamicsConsensus}). Setting the model's mean field consensus time equal to the VM consensus time implies that $\beta_c = \langle k\rangle^{-1}$ is the critical value between CCIS and VM dynamics \footnote{We should mention that more accurate methods for determining the mean consensus time exist for the VM, as explained further in \cite{PAHomo,PAHetero}.}. Neutral spreading (not shown), on the other hand, breaks with the other spreading methods by only depending on the first degree moment, and is therefore mostly independent of the network's degree distribution in the mean field.

Finally, we check whether $T_{\text{cons}}\sim\beta^2$ for each process (Figs. \ref{Tbeta} and \ref{VMDynamicsConsensus}). Agreement with theory is closest when $\beta \sim O(1)$ and $\langle k\rangle\sim 10-20$.  When $\langle k \rangle$ approaches 1 or $\beta\ll1$ we see that $T_\text{cons} \sim \beta^{-1}$. The reason is as follows: the number of nodes convinced at each timestep in this limit is very low (i.e., 2 with probability $\beta^2\approx 0$, 1 with probability $\beta$, and 0 with probability $1-\beta$), therefore, the time until a given node is convinced is a geometric process:

\begin{equation}
p(t) = \beta (1-\beta)^{t-1},
\end{equation}
which would imply the average time until a node is convinced is $\beta^{-1}$. The consensus time would scale similarly.


\bibliographystyle{apsrev4-1}
\bibliography{CompetingContagionsBib.bib}

\end{document}